\documentclass[12pt]{article}

\pdfoutput=1

\usepackage{graphicx}
\usepackage{epstopdf}
\usepackage[centertags]{amsmath}
\usepackage{amsthm}
\usepackage{amsfonts}
\usepackage{ccaption}
\usepackage[usenames]{color}
\usepackage{mathrsfs}
\usepackage[
      colorlinks=true,
      linkcolor=blue,
      urlcolor=blue,
      filecolor=blue,
      citecolor=red,
      pdfstartview=FitV,
      pdftitle={},
        pdfauthor={Veronika Hubeny, Don Marolf, Mukund Rangamani},
        pdfsubject={Black holes on the brane},
        pdfkeywords={black holes, Hawking radiation, deconfinement},
        pdfpagemode=None,
        bookmarksopen=true
      ]{hyperref}

\usepackage{rotating}
\makeatletter
\@addtoreset{equation}{section}

\makeatletter
\renewcommand\section{\@startsection {section}{1}{\z@}%
                                   {-3.5ex \@plus -1ex \@minus -.2ex}%nn
                                   {2.3ex \@plus.2ex}%
                                   {\normalfont\large\bfseries}}
\renewcommand\subsection{\@startsection{subsection}{2}{\z@}%
                                     {-3.25ex\@plus -1ex \@minus -.2ex}%
                                     {1.5ex \@plus .2ex}%
                                     {\normalfont\bfseries}}

  \captionnamefont{\bfseries}
  \captiontitlefont{\small\sffamily}
  \captiondelim{: }
  \hangcaption

\newcommand{\be}{\begin{equation}}
\newcommand{\ee}{\end{equation}}
\newcommand{\beq}{\begin{eqnarray}}
\newcommand{\eeq}{\end{eqnarray}}

\def\sec#1{\S \ref{#1}}
\def\fig#1{Fig.\,\ref{#1}}
\def\req#1{(\ref{#1})}
\def\App#1{Appendix \ref{#1}}

\def\({\left(}
\def\){\right)}
\def\[{\left[}
\def\]{\right]}

\def\ord#1{\CO\left(#1\right)}

\def\p{\partial}

\def\CB{{\cal B}}

\def\CE{{\cal E}}

\def\CM{{\cal M}}
\def\CN{{\cal N}}
\def\CO{{\cal O}}

\def\CS{{\cal S}}

\def\R{{\bf R}}
\def\Sp{{\bf S}}

\def\A5S5{{\rm AdS}_5 \times \S^5}

\def\p{\partial}

\def\ord#1{\CO\!\left( #1 \right)}

\def\AdS#1{AdS$_{#1}$}
\def\SAdS#1{Schwarschild-AdS$_{#1}$}

\def\Uz{U_{,z}}

\def\Vz{V_{,z}}

\def\Zz{Z_{,z}}

\def\H{{\bf H}}
\def\R{{\mathbb R}}

\def\Ff#1{^{#1}{\mathfrak f}}

\title{{\bf \Large Hawking radiation in large N strongly-coupled \\  field theories}}

\author{\normalsize
Veronika E. Hubeny$^{a,b}$\footnote{veronika.hubeny@durham.ac.uk},\ \ Donald Marolf$^{\,c\,}$\footnote{marolf@physics.ucsb.edu}, and
Mukund Rangamani$^{a,b}$\footnote{mukund.rangamani@durham.ac.uk} \\ 
$^a$\small \sl  Centre for Particle Theory \& Department of
Mathematical Sciences,\\[-1.5mm]
\small \sl Science Laboratories, South Road, Durham DH1 3LE, United Kingdom. \\[1mm]
$^b$\small\sl Kavli Institute for Theoretical Physics, UCSB, Santa Barbara, CA 93015, USA.\\[1mm]
$^c$ \small \sl Physics Department, UCSB, Santa Barbara, CA 93106, USA.
}
\begin{document}

\setlength{\baselineskip}{14pt}
\begin{titlepage}
\maketitle

\begin{picture}(0,0)(0,0)
\put(350, 320){DCPT-09/57}
\put(350, 305){NSF-KITP-09-164}
\end{picture}
\vspace{-42pt}

%Abstract
\begin{abstract}
We consider strongly coupled field theories at large $N$ on black hole backgrounds. At sufficiently high Hawking temperature $T_H$, one expects a phase where the black hole is in equilibrium with a deconfined plasma.  We explore this phase in the context of the AdS/CFT correspondence, and argue that two possible behaviors may result.  At a given Hawking temperature $T_H$, field theories on large  black hole backgrounds will result in a plasma that interacts strongly with the black hole.  Such cases will be dual to novel bulk spacetimes having a single connected horizon which we dub {\it black funnels}.   We construct examples of black funnels in low spacetime dimensions for different classes of field theory black holes.   In this case, perturbing the equilibrium state results in the field theory exchanging heat with the black hole at a rate typical of conduction through deconfined plasma. In contrast, we argue that due to the finite physical size of plasma excitations, smaller black holes will couple only weakly to the field theory excitations.  This situation is dual to bulk solutions containing two disconnected horizons which remain to be constructed.  Here perturbations lead to heat exchange at a level typical of confined phases, even when $T_H$ remains far above any deconfinement transition.     At least at large $N$ and strong coupling, these two behaviors are separated by a sharp transition.   Our results also suggest a richer class of brane-world black holes than hitherto anticipated.

\end{abstract}
\thispagestyle{empty}
\setcounter{page}{0}
\end{titlepage}

\renewcommand{\thefootnote}{\arabic{footnote}}
%______________________________________

%%%%%%%%%%%%%%%%%%%%%%%%%%%%%%%%%%%%%%%%%%%%

\tableofcontents

%~~~~~~~~~~~~~~~~~~~~~~~~~~~~~~~~~~~~~~~~~~~~~~~
\section{Introduction}
\label{s:intro}
%~~~~~~~~~~~~~~~~~~~~~~~~~~~~~~~~~~~~~~~~~~~~~~

Studies of quantum field theories in curved spacetimes over the last few decades  have revealed an intriguing variety of physical phenomena such as vacuum polarization, particle production etc., see e.g.\ \cite{Birrell:1982ix,Jacobson:2003vx,Ross:2005sc}  for reviews.  In particular, the discovery of Hawking radiation by black holes has led to deep and fundamental issues, which have played an important role in attempts to formulate a quantum theory of gravity. However, much of this discussion is rooted in the study of free field theories in curved spacetimes. The details, and even what one might call the basic physical processes, are much less clear when the field theory interacts strongly.  Gaining some insight into the strong coupling case will be the primary focus of the current work.

Let us begin by reviewing the situation for free fields on a fixed black hole spacetime.    As is well known,
the black hole radiates and absorbs field theory quanta in a manner much like an ideal black body at the Hawking temperature $T_H$, with absorption and emission coefficients modulated slightly by  grey-body factors which depend on the details of the black hole geometry \cite{Hawking:1974sw,FH}.  Several states for the field theory are commonly discussed.  In particular, the so-called Hartle-Hawking vacuum describes radiation in equilibrium with the black hole, so that the ingoing and outgoing fluxes are balanced.  The renormalized stress tensor in this state is regular on both future and past horizons, and the state is conveniently defined by a Euclidean path integral periodic in imaginary time.  However,
one can also consider the time-asymmetric Unruh vacuum which describes only the outgoing flux of Hawking quanta from the black hole, or one may directly study the  radiation emanating from a black hole formed in a gravitational collapse process.  This radiation rapidly approaches precisely what one would expect from the above grey-body description at temperature $T_H$ \cite{Hawking:1974sw,FH}.  In any of the above settings, it is straightforward to compute occupation numbers and the expected stress tensor far from the black hole, though this becomes more difficult close to the horizon.  If we now allow the  field theory quanta to interact weakly, one expects this basic picture to hold at least in perturbation theory.

In contrast,  for strongly interacting theories it is a challenge to obtain even a qualitative understanding of Hawking radiation.   One can still formally define the Hartle-Hawking state via a Euclidean path integral and, due to periodicity in Euclidean time, it will satisfy the Kubo-Martin-Schwinger (KMS) condition.  In this sense the state is thermal; see e.g. \cite{Jacobson:1995ak,Ross:2005sc} for reviews and references.   In addition, since this construction reduces to the thermal path integral in flat space in the region far from the black hole, the stress tensor in this region will again be thermal.  However, more detailed information is sorely lacking.  In particular one has little knowledge of  the corresponding Unruh states or other out-of-equilibrium situations.  See e.g., \cite{Carter:1976di,FH,MacGibbon:1990zk,MacGibbon:1991tj} for further comments.

One tool that can be used to probe field theories at strong coupling is the Anti-de Sitter/conformal field theory (AdS/CFT) correspondence \cite{Maldacena:1997re,Gubser:1998bc,Witten:1998qj}.
This correspondence maps the dynamics of the field theory to string theory (or classical gravity) in a higher dimensional Anti-de Sitter (AdS) spacetime. The field theories in question are typically non-abelian gauge theories; denoting the rank of the gauge group by $N$ and the 't Hooft coupling by $\lambda$, in the limit of large rank and strong 't Hooft coupling  ($N, \lambda \gg 1$)  the dual description of the field theory reduces to classical gravitational dynamics in the higher dimensional AdS  spacetime.\footnote{For example, in the case of ${\cal N} = 4$ super Yang-Mills dual to AdS${}_5 \times \Sp^5$, the 't Hooft coupling $\lambda$ controls the curvature radius of the AdS spacetime in string units, while the string interactions are governed by $g_s \sim \frac{1}{N}$. Classical gravity is a good description when string interactions are suppressed and the curvatures are small. More generally, the field theory degrees of freedom are measured by a central charge $c$ and for the field theories with known holographic duals $c \sim N^\alpha$ for some $\alpha >0$.} Following standard convention, we refer to AdS/CFT as a holographic duality due to this difference in dimensions and refer to the field theory as living on the conformal boundary of AdS.

We should be clear that Unruh-like and other out-of-equlibrium states remain challenging to describe even in the context of AdS/CFT and may require significant numerical work.  In this paper, we restrict explicit constructions to configurations which, at least in the large $N$, strong coupling limit, describe equilibrium states.   These states are Hartle-Hawking-like in the sense that there is no net transport of heat in this limit.  Because Hartle-Hawking states do not exist for rotating asymptotically flat black holes \cite{Kay:1988mu}, we focus on non-rotating black holes below.  The rotating case will be discussed elsewhere \cite{toappear}.  However, as the reader will see, the dual AdS description of these Hartle-Hawking-like states strongly suggests that novel behaviors arise for nearby out-of-equilibrium states.

Interestingly, as we review in \sec{heuristic}, studies of strongly coupled Hawking radiation performed to date via AdS/CFT suggest that the effects of Hawking radiation may be {\em parametrically} smaller as a function of $N$ at strong coupling than in free theories with the same field content.  Naively one would expect the energy density of Hawking radiation to scale with some positive power of $N$, since the black hole can radiate any of the microscopic field theory degrees of freedom with equal probability. However, an explicit analysis using the bulk gravitational description reveals an $\ord{1}$ energy density. This conclusion is arrived at by studying known black hole solutions in AdS spacetimes which asymptote to the given black hole metric on the boundary of AdS, as this is the black hole background on which the field theory lives   \cite{Chamblin:1999by,Astafanesi:2004bh,Fitzpatrick:2006cd,Gregory:2008br}.
  A partial explanation of this feature was offered in \cite{Fitzpatrick:2006cd}, which noted that the field theories under study exhibit confined phases\footnote{Our focus will be on large $N$ gauge theories which do not exhibit a dynamical mass gap (unlike QCD). The notion of confinement versus deconfinement is simply the statement about the free energy;  it is $\ord{1}$ in the confined and $\ord{N^2}$ in the deconfined phases. } where, even at large $N$, the density of propagating modes remains only ${\cal O}(1)$ and does not grow with $N$.  In such a phase, Hawking radiation far from the black hole would be correspondingly reduced.

However, the field theories also exhibit deconfined phases, where the density of propagating modes is much larger (of $\ord{N^\alpha}$ for some $\alpha > 0$).  Furthermore, at high temperatures the confined phase is thermodynamically unstable or, at best, meta-stable and decays to the deconfined phase.  It therefore remains to understand Hawking radiation in the deconfined phase, as well as transitions between the two phases in the presence of black holes.  For example, to what extent does a hot black hole tend to ``melt" the confined phase and to catalyze a transition to a deconfined phase?  These are the questions  we begin to explore below.

The previous literature on this subject has largely focussed on the case where the large $N$ field theory is coupled to dynamical gravity.  This situation is holographically dual \cite{Gubser:1999vj} to Randall-Sundrum braneworld models  \cite{Randall:1999vf}, and much of the interest centered on possible implications for brane-world phenomenology.  In particular, it was argued in \cite{Tanaka:2002rb,Emparan:2002px} that considerations of Hawking radiation in the large $N$ theory implied that black holes in such a brane-world setting would rapidly disappear from the brane.   The point here was that, if free field theory could be used as a guide, the timescale for black hole evaporation would be a certain function of $N$ and the black hole mass ($M$) times the effective Newton constant ($G_{brane}$) on the brane, $M \, G_{brane}$, which, under the holographic duality, maps to a classical timescale in the bulk gravitational description.  As a result, even at the purely classical level in the bulk, refs.\ \cite{Tanaka:2002rb,Emparan:2002px}  argued that there should be no static black hole solutions in these braneworld scenarios.
While the above argument was later countered by the observation of \cite{Fitzpatrick:2006cd} that the large $N$ theory might be in a confined phase, many issues remain open; see e.g.  \cite{Chamblin:1999by,Shiromizu:2000pg,Kudoh:2003xz,Kudoh:2004kf,Tanaka:2007xm,Tanahashi:2007wt,Yoshino:2008rx,Harvey} for summaries and recent progress.

Although the braneworld questions continue to be of interest, we find it useful at both the conceptual and practical level to focus on the limit in which gravity on the brane decouples and the field theory propagates on a fixed non-dynamical black hole background.  From the bulk perspective, this is the limit where the brane recedes to the AdS boundary  and the metric on the brane becomes simply a boundary condition that one imposes by hand.   We focus on this limiting case throughout most of this work, though we mention implications for dynamical braneworld black holes in \sec{s:discuss}.

We begin by using AdS/CFT to give a heuristic picture of Hawking radiation for strongly coupled large $N$ field theories in \sec{heuristic}, where we also review the relevant known results.  In a deconfined phase, we argue that two possible behaviors may result, depending on the size $R$ and the Hawking temperature $T_H$ of the black hole in relation to various scales in the field theory.
Note that since our field theory lives on a fixed background spacetime in which gravity is non-dynamical, we are free to consider arbitrary black hole geometries which need not satisfy any equations of motion.   As a result, we may regard $R$ and $T_H$ as completely independent quantities.     In terms of more familiar examples, one may think of Reisner-Nordstrom  black holes where one may adjust the relative size of $R$ and $T_H$ by changing the charge.
We refer to these arbitrarily specified black holes as either {\em field theory black holes} or {\em boundary black holes}  to distinguish them from {\em bulk} black holes that arise in the dual AdS description; the latter will satisfy Einstein's equation with a cosmological constant as well as a boundary condition determined by the field theory black hole metric.

At given $T_H$, we will argue that large asymptotically flat field theory black holes couple strongly to the deconfined plasma.  Here we use the term coupling in a loose sense that characterizes  how excitations around the Hartle-Hawking state interact with the (non-dynamical) black hole.  Strong coupling means that deviations from the Hartle-Hawking state relax back to equilibrium on a timescale set by the black hole size, while weak coupling means that such deviations decay much more slowly.

\begin{figure}[h]
\begin{center}
\includegraphics[scale=1.1]{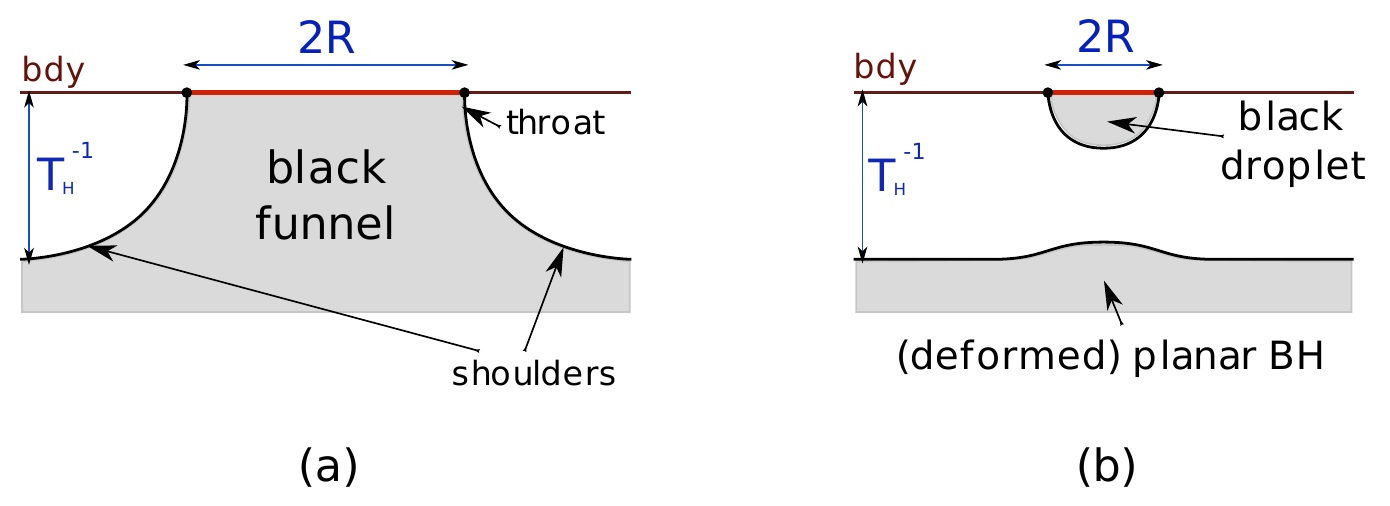}
 \caption{A sketch of our two novel classes of solutions: {\bf (a):} black funnel and  {\bf (b):} black droplet above a deformed planar black hole.  In both cases the boundary metric is that of a black hole of temperature $T_H$ and horizon size $R$ (the top line corresponds to the boundary, the dots denote the horizon of the boundary black hole). Here we fix $T_H$ and vary $R$; (a) for sufficiently large $R$, we expect the bulk solution to have a single connected horizon in a ``funnel" shape, while (b) for small $R$, we expect two disconnected horizons, a ``droplet" and a planar black hole.  The shaded regions are those inside the bulk horizons.}		
\label{f:fundrop}
\end{center}
\end{figure}

Consider first strongly coupled large field theory black holes as defined above. For these the bulk AdS duals  of the corresponding Hartle-Hawking states will be novel spacetimes having a single connected horizon which we dub {\it black funnels} (see \fig{f:fundrop}a).   In this case, performing work on the system or introducing additional heat sources or sinks will result in heat exchange with the field theory black hole at a rate typical of heat flow through deconfined plasma.  In the bulk, this heat will simply flow along the horizon from the boundary black hole down the throat of the funnel and out along the shoulders.

In contrast, we argue that the Hartle-Hawking states of smaller asymptotically flat black holes will be dual to bulk AdS solutions containing two disconnected horizons.  One bulk horizon reaches the AdS boundary, where it ends on the horizon of the boundary black hole.  We refer to this bulk horizon as a {\em black droplet}, as we expect this horizon to cap off smoothly in the bulk (see \fig{f:fundrop}b).  The other horizon will asymptotically approach that of a planar black hole\footnote{For asymptotically flat boundary metrics.  We will also consider boundaries which are asymptotically $H^n \times {\mathbb R}$, in which case this bulk horizon approaches that of the  black holes discussed in \cite{Emparan:1999gf}. }.  In a slight abuse of language, it will be convenient to overlook this distortion and continue to refer to this horizon as a planar black hole.

% Figure

The lack of connection between the two bulk horizons for small boundary black holes suggests that performing work on the system or introducing additional heat sources or sinks results in rather little heat exchange between the two bulk black holes, and thus rather little exchange between the field theory black hole and the field theory plasma.     As a result, the amount of heat and other quantities transported outward from the field theory black hole will be similar to what one would expect in a confined phase, even if $T_H$ is far above the temperature $T_D$ of any deconfinement transition. From the field theory point of view, we interpret the weak coupling between the black hole and plasma as being due to the finite (and large) physical size $R_e$ of plasma excitations relative to the black hole, $R_e \gg R$.

Following this heuristic discussion, we construct examples of black funnels in low spacetime dimensions for various classes of boundary black holes.  In \sec{s:bfunnelbtz} we consider asymptotically flat 1+1 field theory black holes.  As we discuss below, all 1+1 black holes should be thought of as having a large size $R$ in comparison with any other scale.  As a result, only black funnel solutions will be expected from the arguments of \sec{heuristic} and this is indeed what we find.    

We then turn to higher dimensional examples; \sec{s:adscmet} considers 2+1 black holes on the boundary of asymptotically \AdS{4} spacetimes exploiting the known AdS C-metric solutions.   Here our particular method of constructing the solution generates  2+1 boundary black holes which are asymptotic to $\H^2 \times {\mathbb R}$  where $\H^2$ is the (Riemannian-signature) hyperbolic plane.  In this context we find both black funnels and black droplets suspended above disconnected horizons. We conclude with a summary and a discussion of the implications for braneworld black holes in \sec{s:discuss}.

%~~~~~~~~~~~~~~~~~~~~~~~~~~~~~~~~~~~~~~~~~~~~~~~
\section{Phases of CFTs on black hole backgrounds}
\label{heuristic}

%~~~~~~~~~~~~~~~~~~~~~~~~~~~~~~~~~~~~~~~~~~~~~~

As described in the Introduction, we wish to investigate strongly interacting quantum fields in a black hole spacetime.   To this end we will employ the AdS/CFT correspondence in the planar limit.   Thus, at least for the moment, we restrict attention to field theories for which this correspondence applies.  Our goal, however, is to extract certain general lessons which we hope will be applicable to a much wider set of models.

The essence of the AdS/CFT correspondence is that the strongly coupled field theory dynamics is recorded in terms of string theory (or classical gravity if the field theory admits an appropriate planar limit) with appropriate asymptotically Anti-de Sitter boundary conditions. If the bulk AdS spacetime geometry is some negatively curved $(d+1)$-dimensional Lorentzian manifold, $\CM_{d+1}$, with conformal boundary $\p \CM_{d}$, then the  field theory lives on a spacetime $\CB_d$ of dimension $d$ in the same conformal class as $\p \CM_{d}$. Choosing an appropriate conformal frame, one may identify $\CB_d$ and $\p \CM_{d}$ and speak of the field theory as living on the AdS boundary.  From the standpoint of the bulk theory, the choice of metric on $\CB_d$ fixes a boundary condition that the bulk solution must satisfy.\footnote{In standard AdS/CFT parlance, this amounts to fixing the non-normalizable mode of the bulk graviton to obtain the desired metric on $\CB_d$.} The correspondence is simplest to state for conformal field theories in dimensions where the trace anomaly vanishes, but with appropriate care the correspondence also holds in the presence of a trace anomaly and it can accommodate non-conformal deformations.

Given this set-up we can probe the details of Hawking radiation in the strongly coupled field theory. More interestingly, as described in \sec{s:intro}, if the field theory under consideration exhibits non-trivial phase structure  we should be able to ascertain the holographic duals of each distinct phase.    To consider field theory in a black hole background, we merely choose $\CB_d$ to be a black hole spacetime.  The space of field theory states on $\CB_d$ then corresponds to the space of asymptotically AdS states of string theory subject to the  boundary condition above.  In the planar limit of the field theory, the latter is essentially the solution space of classical (super-)gravity with the given boundary condition.  The different phases of the CFT on $\CB_d$ should thus correspond to different bulk spacetime geometries $\CM_{d+1}$ in this solution space.

We work in a universal sector of the AdS/CFT correspondence, wherein the bulk spacetime $\CM_{d+1}$ is a solution to Einstein's equations with a negative cosmological constant. The bulk action is therefore taken to be\footnote{We will use $g_{AB}$ to denote the bulk metric on $\CM_{d+1}$ with uppercase Latin indices indicating the bulk  spacetime dimensions. The metric on $\CB_d$ will be denoted as  $\gamma_{\mu \nu}$ with lowercase Greek indices labeling boundary directions.}
\begin{equation}
\CS_{\text{bulk}}  = \frac{1}{16 \pi \, G^{(d+1)}_N}\,\int d^{d+1}x \, \sqrt{-g}\, \left( R - 2 \, \Lambda\right)  \ ,
\label{bulkact}
\end{equation}	
where we can express the cosmological constant as $\Lambda = -\frac{d\,(d-1)}{2}\,\ell_{d+1}^{-2}$ in terms of the AdS$_{d+1}$ radius $\ell_{d+1}$.  From the field theory side we concentrate on the response of the energy-momentum tensor to the non-trivial background on $\CB_d$.

Let us consider the simple case where the boundary spacetime $\CB_d$ has a Killing field $\xi_{\CB}$ which generates the black hole horizon and which is timelike in some asymptotic region far from the black hole.  For simplicity we also take the black hole horizon in $\CB_d$  to have a single connected component, although this is not imposed on us by any equations of motion.  Field theory states that are stationary with respect to $\xi_{\CB}$ are dual to bulk solutions with a Killing field $\xi$ which agrees with $\xi_{\CB}$ on the boundary.  We will focus on equilibrium states so that, as with the Hartle-Hawking vacuum, the field theory stress tensor should be regular on both future and past horizons.   As a result, the bulk spacetime should also have a regular Killing horizon.  The problem at hand thus reduces to finding distinct black hole solutions for a gravitational theory in $d+1$ spacetime dimensions, all of which obey the prescribed boundary conditions.  The phase structure of the CFT on $\CB_d$ is just the classical phase structure of the corresponding black hole spacetimes.

After a brief review of deconfinement transitions in AdS/CFT in \sec{review},
we give a heuristic discussion of the desired bulk black hole spacetimes in \sec{hBH} below.   The main conclusion is that bulk solutions dual to field theory black holes in equilibrium with a deconfined plasma may display two possible behaviors.  The bulk Killing horizon may be connected, a case we refer to as a black funnel, or it may be disconnected.  Which possibility is realized is determined by the size $R$ and the temperature $T_H$ of the boundary black hole in comparison with certain scales set by the field theory.  A sharp phase transition is expected as one tunes the values of $R,T_H$ though the critical values.\footnote{Tuning $R$ and $T_H$ involves changing the metric on the boundary manifold $\CB_d$. We will treat the background spacetime as an external parameter for the field theory and describe the different possibilities as distinct phases of the field theory.}  In \sec{hFT} we interpret our conclusions in field theory language: in the connected case the black hole couples strongly to the deconfined plasma while there is no such coupling in the disconnected case.  In this latter situation it appears that this coupling is prohibited by the  intrinsic size of the field theory excitations which have condensed to form the plasma.    This discussion sets the stage for the rest of the paper, which focusses on constructing such solutions.

%~~~~~~~~~~~~~~~~~~~~~~~~~~~~~~~~~~~~~~~~~~~~~~~~~~~~~~~~~~~
\subsection{Brief Review: Deconfinement transitions in AdS/CFT}
\label{review}
%~~~~~~~~~~~~~~~~~~~~~~~~~~~~~~~~~~~~~~~~~~~~~~~~~~~~~~~~~~~~

It is useful to begin by reviewing the AdS/CFT description of the confinement/deconfinement transition; we will first discuss the case where there is no black hole on the boundary. Moreover, we take $\CB_d$ to have compact spatial sections where there is a sharp transition in the planar (large $N$) limit. For instance, one can consider a CFT on the Einstein Static Universe, i.e., $\CB_d = \R \times \Sp^{d-1}$, which is the boundary of the global AdS spacetime. As a prototype example, the reader can keep in  mind the phase structure of the four dimensional $\CN =4$ Super-Yang Mills (SYM) in $d=4$, but the discussion extends analogously to other dimensions.

At low temperature we have the confined phase with $\ord{1}$ free energy and at high temperatures a deconfined phase with $\ord{c}$ free energy, where $c$ is the central charge\footnote{The central charge $c$ can be defined either from the two point function of the energy momentum tensor, or by looking at the entropy density of a CFT at finite temperature, which from dimensional analysis scales as $s\sim c\, T^{d-1}$.} of the CFT which we take to scale as $c \sim N^\alpha$. In the case of $\CN =4$ SYM, $\alpha =2$, reflecting the non-abelian gluon degrees of freedom. In the holographic dual this transition is the Hawking-Page transition \cite{Hawking:1982dh} where the thermal \AdS{d+1} geometry exchanges dominance with the \SAdS{d+1} spacetime \cite{Witten:1998zw}.\footnote{A similar confinement-deconfinement transition can be seen at weak coupling as well  for four dimensional large $N$ gauge theories on compact spatial manifolds \cite{Sundborg:1999ue,Aharony:2003sx}. See also \cite{Gaiotto:2007qi} for analogous results for three dimensional Chern-Simons matter theories.} The phase transition occurs at the deconfinement temperature $T_D$ which is set by the curvature radius of the $\Sp^{d-1}$ (which is the only other dimensionful parameter in the problem if we consider conformal field theories).

One can also consider the situation where the boundary is just a Minkowski space, $\CB_d = \R^{d-1,1}$, which is of greater relevance to our immediate considerations.
In this case there isn't a phase transition since the only scale in the problem is the temperature. As we will describe below, at any non-zero temperature the field theory always prefers to live in the deconfined phase. Formally, however, one can consider the two distinct phases and their dual geometries to gain intuition for the physics of the field theories. The relevant bulk solutions are i) pure \AdS{d+1} spacetime expressed in a conformal frame such that the boundary metric just the flat Minkowski metric, e.g.,
\begin{equation}
ds^2_{\text{planar AdS}} = \frac{1}{z^2}\, \left(-dt^2 + d{\bf x}_{d-1}^2  + dz^2  \right)  \ ,
\label{planarAdS}
\end{equation}	
which is dual to the confined phase, and ii) the planar \SAdS{d+1} black hole,
\begin{equation}
ds^2_\text{planar BH} = \frac{1}{z^2}\, \left(- \Ff{d}(z) \, dt^2 + d{\bf x}_{d-1}^2    + \frac{dz^2}{\Ff{d}(z)} \right) \ ,
\label{planarBH}
\end{equation}	
dual to the deconfined phase.
Here and below we use $t,x^i$ to denote coordinates along the boundary at $z=0$.  In this sense, $z$ is a radial coordinate in AdS.  In addition we have
\begin{equation}
\Ff{d}(z) = 1 -\frac{z^{d}}{z_0^{d}} \ ,
\label{Ff0def}
\end{equation}	
so that the horizon is located at $z=z_0$ and has temperature $T = \frac{d}{4\pi\, z_0}$.  The stress tensor of the large $N$ field theory is given by the boundary stress tensor (aka ``holographic stress tensor'') of the bulk solution \cite{Henningson:1998gx,Balasubramanian:1999re}, which is readily computed (following, say, \cite{deHaro:2000xn}) from the Fefferman-Graham expansion \cite{FG}; i.e., by expanding the metric in powers of $z$ in the above coordinates.    One finds (see for example \cite{Bhattacharyya:2008jc}):
\begin{equation}
T^\text{planar  AdS}_{\mu\nu}  = 0 ,  \qquad
T^\text{planar  BH}_{\mu\nu} = \frac{(\pi T)^4}{16 \pi G_N^{(d+1)}}\, \left(\eta^{\mu\nu} + 4 \, \delta^\mu_t \, \delta^\nu_t\right).
\end{equation}
In particular, we should note that the correspondence relates  the central charge  of the field theory to the AdS radius measured in Planck units, i.e., $c = \frac{\ell_{d+1}^{d-1}}{16\, \pi\, G_N^{(d+1)}}$, providing the desired mapping between the field theory and the gravitational parameters. In the case of $\CN =4$ SYM in four dimensions we have $c = N^2 =  \frac{\pi}{2 \pi G_N^{(5)}}$. The planar black hole corresponds to a field theory state with stress energy of order $c$, consistent with the interpretation as a deconfined phase.

Comparing the free energies of the corresponding Euclidean spacetimes, one finds that  the planar black hole has lower free energy for any $T > 0$.  In fact, for $T > 0$ the confined phase is thermodynamically unstable.  This can be seen from the fact that the horizon of (\ref{planarAdS}) is degenerate.  As a result, while the thermal circle in the Euclidean solution shrinks to zero size near the horizon, it remains non-contractible.  This means that strings wrapped on the thermal circle have a tachyon near the horizon \cite{Horowitz:1996qd}, rendering the phase unstable.\footnote{A similar instability would be expected in any theory of quantum gravity that allows topology change, as reducing the size of the infinite throat tends to decrease the action.}  In contrast, the planar black hole is thermodynamically stable.

%~~~~~~~~~~~~~~~~~~~~~~~~~~~~~~~~~~~~~~~~~~~~~~~~~~~~~~~~~~~~
\subsection{Boundary black holes and deconfined plasmas: the bulk description}
\label{hBH}
%~~~~~~~~~~~~~~~~~~~~~~~~~~~~~~~~~~~~~~~~~~~~~~~~~~~~~~~~~~~~

It is an interesting fact that,  for the only known asymptotically locally AdS solutions with asymptotically flat boundary black holes,  the boundary stress tensor vanishes far from the black hole \cite{Astafanesi:2004bh,Fitzpatrick:2006cd}.  These solutions are essentially the Schwarzschild black string solution of \cite{Chamblin:1999by} up to minor modifications.\footnote{Ref. \cite{Fitzpatrick:2006cd} cuts off the singular part of the solution with a so-called IR brane.   Perhaps a nicer modification is to replace the boundary metric with a Schwarzschild black string compactified on a Scherk-Schwarz circle so that the spacetime develops a smooth IR floor.  See \sec{s:discuss} for comments on this situation. }  Numerical investigations to date \cite{Kudoh:2004kf,Tanahashi:2007wt} also suggest solutions of this sort; see however \cite{Yoshino:2008rx}.\footnote{Here the numerical investigations have been carried out for the case of braneworld models with a UV brane located at a finite radial scale in the bulk AdS spacetime so that gravity on the brane is also dynamical.}
As a result, these solutions appear to represent a confined phase of the dual field theory in the black hole background.

In contrast,  we wish to study Hartle-Hawking-like states of the field theory on black hole spacetimes.  For concreteness, we suppose that $\CB_d$ is asymptotically flat and contains a single connected horizon.  Other cases where  $\xi_{\CB}$ becomes asymptotically constant are similar, though the AdS-like case where the norm grows asymtotically is more subtle and will be discussed elsewhere \cite{toappear}.  As noted in \sec{s:intro}, for the asymptotically flat case the Euclidean path integral tells us that $T_{\mu \nu}$ should approach the stress tensor of a thermal fluid in the asymptotic region.  As a result, the \AdS{d+1} bulk dual should approach the planar AdS black hole (\ref{planarBH}) at large $x^i$, with a corresponding horizon localized near $z = \frac{d}{4\pi \,T_H}$.   Furthermore, in any  region near the \AdS{d+1} boundary ($z=0$), there must be a horizon, whose restriction to $\CB_d$ coincides with that of the boundary black hole.

There are clearly two possibilities:  either these two horizons join to form a single connected horizon, or they are disconnected, see \fig{f:fundrop}. We now argue that which possibility occurs is primarily determined by the size $R$ of the boundary black hole and its temperature $T_H$.  In particular, we expect disconnected horizons for $R \, T_H \ll 1$, but connected horizons for $R \, T_H \gg 1$, though the precise value of $R \, T_H$ at the transition may depend on the details of both the black hole geometry $\CB_d$ and the particular field theory in question.

The basic argument is quite simple.  Consider the planar black hole (\ref{planarBH}) at some temperature $T = T_H$.  This solution is stable to small perturbations.  We now wish to perturb this solution by adding a black hole of the same temperature $T_H$ to the boundary metric.  Our general experience with AdS spaces suggests that the effect of a disturbance of size $R$ on the boundary falls off deep in the bulk and typically extends to $z \sim R$.  This phenomenon is seen in perturbative studies\footnote{For instance one can consider a Gaussian source $\phi_0(x) = e^{-\frac{x^2}{2 R^2}}$ for some gauge invariant operator on the boundary. In the bulk this source will produce a field $\phi(z,x)$ which can be determined using the bulk to boundary propagator \cite{Witten:1998qj}. A simple calculation of convolving the Gaussian waveform with the bulk to boundary propagator reveals that the bulk field penetrates down to  $z \sim R$.} or, in perhaps a closer parallel to black hole horizons, studies of minimal surfaces which intersect the boundary in spheres of size $R$, see e.g., \cite{Mateos:2006nu}.   Thus, since the planar black hole horizon is at $z = \frac{d}{4\pi\, T_H}$, the effect on this horizon of adding the black hole to the boundary should be perturbative for $R \, T_H\ll \frac{d}{4\pi}$.  In this case we expect to find a static bulk solution with two disconnected horizons: one horizon will be a perturbed planar black hole, while the other should hang down into the bulk from the boundary black hole, with the horizon capping off smoothly far above the (perturbed) planar horizon.  We refer to the first horizon as simply the planar black hole, while the second horizon we christen a {\it black droplet.}  See \fig{f:fundrop}b for a sketch of the solution.

In contrast, for $R\, T_H \gg \frac{d}{4 \pi}$ the horizon of the planar black hole will be strongly perturbed.  If one imagined a two-horizon solution containing a black droplet and a separate planar black hole, the horizons would nearly touch.  It is natural to assume that the horizons merge, resulting in a single connected horizon which we call a {\em black funnel}. It is an interesting question whether black funnels also exist for $R \,T_H\ll \frac{d}{4\pi}$ and, if so, how their free energies compare to the corresponding two horizon (droplet + planar black hole) solution.

We note that the above argument can be generalized in several ways.  One could consider non-equilibrium solutions in which the horizons have different temperatures.  So long as the horizons remain disconnected, this need not be a problem in classical gravity; i.e.\ in the limit $N, \lambda \gg 1$.  Our argument suggests that only the size $R$ of the boundary black hole and the temperature $T_{planar}$ of the planar black hole are relevant, and that the horizons should again remain disconnected for $R \, T_{planar} \ll d/4\pi$, and in particular in the limit $T_{planar} \rightarrow 0$, in which the planar black hole horizon becomes just the Poincar\'e horizon of planar AdS.   For $R \, T_{planar} \gg d/4\pi$ we again expect the horizons to join.  However, from the field theory viewpoint, any stationary non-equilibrium situation must have a constant flow of heat.  Such terms in the boundary stress tensor necessarily diverge on either the past or the future horizon, as is familiar from Unruh states. Thus, only in equilibrium states can the stress tensor be regular on both horizons.  As a result, a bulk black funnel with a smooth bifurcate Killing horizon should exist only  when the temperature $T_H$ of the boundary black hole satisfies $T_H =  T_{planar}$.

The above discussion raises many interesting questions concerning both non-equilibrium black holes and the detailed phase diagram of equilibrium AdS black holes when the boundary metric $\CB_d$ is a black hole spacetime.   However, since these issues are a departure from our main focus, we save such discussion for \sec{s:discuss}.

%~~~~~~~~~~~~~~~~~~~~~~~~~~~~~~~~~~~~~~~~~~~~~~~~~~~~~~~~~~~~~
\subsection{Field theory interpretation:  excitations of finite size}
\label{hFT}
%~~~~~~~~~~~~~~~~~~~~~~~~~~~~~~~~~~~~~~~~~~~~~~~~~~~~~~~~~~~~~~~

It is natural to ask what each class of AdS solutions suggested above would imply for the dual quantum field theory.  The black funnels are easy to interpret.  On the field theory side, they represent black holes in equilibrium with a deconfined plasma with which they interact strongly.  In particular, the bulk horizon will conduct heat, charge, etc to the boundary black hole at a rate determined by an appropriate (positive) power of $N$.  From the field theory perspective, the state resembles  a large-$N$ analogue of the free Hartle-Hawking state, with the energy density scaled up by an appropriate power of $N$.  In particular, consider such a state at finite volume (e.g., on a torus).  If the volume is either increased or decreased by a small amount, one expects heat to flow either out of or into the black hole so that the solution quickly settles down to a new Hartle-Hawking-like state at the same temperature.  More dramatic non-equilibrium effects may also be possible and will be discussed in \sec{s:discuss}.

In contrast, the two-horizon solutions present a greater surprise.  They represent black holes in equilibrium with a deconfined plasma to which they are only very weakly coupled in the following sense: One notes that the plasma excitations, thought of as small deformations of the bulk planar black hole, can pass directly under the black droplet with minimal interaction. The two horizons can exchange heat only via the $\ord{1}$ fields that propagate in the bulk.  For example, suppose we have such a state for a field theory in finite volume and then either increase or decrease this volume by a small amount.  At the classical level, the bulk AdS solution clearly responds by increasing or decreasing the height of the planar black hole horizon; i.e., by heating or cooling the plasma so that its temperature $T$  no longer matches $T_H$ of the boundary black hole.  It is only over very long time scales that quantum effects in the bulk (which are suppressed by a factor of $G_N^{(d+1)}T_H^{d-1}$) will return the plasma to the black hole temperature $T_H$.  This slow rate of heat exchange with the field theory black hole is more typical of what one would expect in a {\it confined} phase.

The reader will note that such behavior is radically different from the Hawking effect for free fields. We should thus give some explanation from the field theory perspective of how this weak coupling might arise.  We believe that the answer is best understood by thinking of the plasma excitations as having a finite physical size $R_e$, which happens to be rather larger than the size $R$ of the black hole.    An analogy might be the absorption and emission of, say, protons by a TeV-scale Schwarschild black hole:\footnote{Or, more picturesquely, of battleships by Schwarzschild black holes with, say,  $M \sim 10^{11}\,\text{kg}$ and $R \sim 1 \, \text{fm}$.} even if the  proton center-of-mass wavefunction is tuned to hit the black hole smack-on, and even if the width of this wavefunction is much smaller than the black hole size, the most likely scenario is that the black hole just encounters empty space between the constituents and that the proton passes right ``around'' the black hole.  There is, of course, a small probability that the constituents will happen to be very close together (and thus close to the center of mass) when they pass the black hole, so that the proton is in fact absorbed.  If, say, a single quark is absorbed, the result will depend on the total momentum of the proton and the fraction of momentum carried by the absorbed quark.  While the proton will be briefly attached to the proton by a QCD string, the string will quickly break if the proton momentum is high enough. The net result will be just a scattering of the proton by the black hole.  See e.g.\ \cite{Marolf:2003wu} for a previous discussion of emission of macroscopic objects by black holes. This is in contrast with the behavior of free fields for which excitations are always effectively point-like and therefore $R_e=0$.

Our picture of the plasma excitations having a large physical size $R_e \gg R$ is analogous to the situation for glueball states in AdS/CFT.  A notion of the physical size of glueballs was discussed in \cite{Polchinski:2001ju}, associated with their distribution of stress energy; i.e., a gravitational form factor.  Glueballs in Minkowski space are represented by gravitational waves in the AdS dual, and gravitational waves at some $z$ in the coordinates of \eqref{planarAdS} corresponded to glueballs of size $\sim z$.  Let us now add a black hole on the boundary.  For a small black hole of size $R$ on the boundary, one expects to find a droplet solution that extends in the bulk to roughly $z \sim R$.  Gravitational waves at $z \gg R$ will thus pass directly under the horizon, as did our plasma perturbations above.\footnote{The weak coupling of such glueballs to the black hole was also noted in \cite{Fitzpatrick:2006cd}.  There it was ascribed to a presumed modification of the glueball spectrum by the black hole background which was not explained in detail.}  The observation that the tail of the graviton's radial wavefunction at the black hole horizon will determine the absorption probability gives an additional argument for thinking of the graviton's radial wavefunction as a wavefunction for the physical size of the glueball.  It would be very interesting to study such scattering and absorption processes in detail.

In summary, we find that both black-funnel and two-horizon scenarios for AdS bulk solutions admit consistent field theory interpretations.  In a given field theory, the result that occurs will be determined by the size $R_e$ of plasma excitations relative to the size $R$ of the black hole at a given temperature $T$.  For field theories with AdS duals, $R_e \sim T^{-1}$ and the relevant criterion is $R \, T \ll 1$ or $R \, T \gg 1$.  The details of the intermediate region ($R \, T \sim 1$; i.e., Schwarschild black holes) would be interesting to explore, as would the dependence on the details of the field theory.  However, in \sec{s:bfunnelbtz} and \sec{s:adscmet} below we simply focus on constructing simple analytic examples of asymptotically AdS spacetimes contains black funnels and black droplets.

%~~~~~~~~~~~~~~~~~~~~~~~~~~~~~~~~~~~~~~~~~~~~~~~
\section{Black funnels in \AdS{3}}
\label{s:bfunnelbtz}
%~~~~~~~~~~~~~~~~~~~~~~~~~~~~~~~~~~~~~~~~~~~~~~

We now turn to the construction of bulk AdS solutions which we expect to be dual to field theory Hartle-Hawking states.
Since both gravity and conformal field theories are known to simplify in low dimensions, we begin with \AdS{3} duals to 1+1 CFTs.  To begin, recall from \cite{Christensen:1977jc} that  given any smooth static asymptotically flat 1+1 black hole spacetime with $T_H \neq 0$,  there is a unique time-translation-invariant stress tensor which is i) smooth on both past and future horizons, ii) conserved in the sense $\nabla_\mu T^{\mu\nu} =0$, and iii) consistent with the trace anomaly for given central charge $c$.  Furthermore, in the asymptotic region $T_{\mu\nu}$ reduces to the thermal stress tensor at the Hawking temperature $T_H$.  This thermal stress tensor is proportional to $c\,T_H^2$, so that it necessarily represents a deconfined phase; no confined phase can be in equilibrium with the black hole.

This uniqueness result can be derived using the fact that any 1+1 spacetime is related to Minkowski space by a conformal transformation, and then using the associated transformation rules for stress tensors.   Similar reasoning can be applied to the boundary stress tensor $T_{\mu \nu}$ of any \AdS{3} solution with a similar conclusion:  for any static smooth \AdS{3} solution whose boundary is a
static asymptotically flat 1+1 dimensional black hole spacetime with $T_H \neq 0$, $T_{\mu \nu}$  must approach that of the planar black hole at large $x^i$.  We note that black droplets are consistent with this result only when accompanied by planar black holes at temperature $T_H$.  But since all solutions of the 2+1  $\Lambda < 0$ Einstein equations are quotients of \AdS{3}, we also know that any Killing horizon is connected.  It follows that droplet/planar black hole pairs cannot arise and we can find only black funnels.
  
This is consistent with the general reasoning in \sec{heuristic}.  Since the bifurcation surface of a 1+1 horizon consists only of a single point, there is no real notion of the size $R$ for such horizons.  However, 1+1 black holes are most closely associated with planar black holes in higher dimensions for which $R = \infty$.    From the field theory point of view, this is because 1+1 horizons can absorb any excitation, no matter how large.  Larger excitations merely take more time to fully cross the horizon.  From the bulk \AdS{3} point of view, this connection is suggested by the analogy between black droplets and  membranes dangling from the AdS boundary.  Since a horizon in \AdS{3} would be modeled by the worldsheet of a string with only one endpoint attached to the AdS boundary, the string will extend infinitely far into the bulk and will intersect the horizon of any planar black hole.  The arguments of \sec{heuristic} thus suggest that Hartle-Hawking states for 1+1 field theory black holes of any $T_H$ are dual to black funnels in \AdS{3}.

Our construction of the funnels is also closely connected to the reasoning in \cite{Christensen:1977jc}.  As noted above, any bulk solution is locally diffeormorphic to \AdS{3}.  Since we wish to study 1+1 black hole spacetimes with topology $\mathbb{R}^2$, we will seek solutions diffeormorphic to
AdS$_3$ in Poincare coordinates,
\begin{equation}
ds^2 = { - dU \, dV + dZ^2 \over Z^2 }.
\label{AdS3}
\end{equation}	
Using the conformal factor $Z$, the boundary metric for \req{AdS3}  is $ds^2_{bdy} = -dU dV$ on ${\mathbb R}^2$.  We wish to find coordinates $u,v,z$ associated with a  change of conformal frame so that the boundary metric becomes that of a static 1+1 black hole.  For definiteness, we choose the boundary metric
\begin{equation}
ds^2_{bndy} =  \({-1 \over 1 - u \, v } \) \, du \, dv  = - \tanh^2 r \, dt^2 + dr^2,
\label{BF3formbdy}
\end{equation}	
where
\begin{equation}
\label{uvrt}
u = e^t \sinh r \ \ \ \ {\rm and}  \ \ \ v = -  e^{-t} \sinh r.
\end{equation}
 Our coordinate transformation must be smooth near the boundary and, in order to simplify the computation of the boundary stress tensor, we assume it to preserve the so-called Fefferman-Graham form \cite{FG}:
\begin{equation}
ds^2 = {1 \over z^2 } \, \[ \({-1 \over 1 - u \, v } +  \CO(z^2) \) \, du \, dv +
 \CO(z^2) (du^2 + dv^2) + dz^2 \].
\label{BF3form}
\end{equation}	
In particular, we require
\begin{eqnarray}
g_{zz} = {\Zz^2-\Uz \, \Vz \over Z^2} = {1 \over z^2}, \ \ \
\label{gzzconstraint}
Z^2 \, g_{uz} =  0, \ \ \
{\rm and \ \ }
Z^2 \, g_{vz} =  0.
\label{gzuconstraint}
\end{eqnarray}	

Since we seek bulk solutions dual to equilibrium states, the Killing field $\partial_t = u \,\partial_u - v \,\partial_v$ of the boundary metric must also Lie derive the boundary stress tensor, which is essentially a trace-reversed version of the $\ord{z^2}$ $u,v$ components of (\ref{BF3form}) \cite{deHaro:2000xn}.   Because the boundary metric and stress tensor determine the full Fefferman-Graham expansion \cite{FG}, it follows that $\partial_t = u \,\partial_u - v \,\partial v$ must be an isometry of the full bulk metric as well.    But all \AdS{3} isometries with a fixed point on the boundary are related by conjugation to  $U \,\partial_U - V \,\partial_V$, so we may choose $\partial_t = u\, \partial_u - v \,\partial v = U\, \partial_U - V \,\partial_V$  without loss of generality.  It follows that $U =  u \,\tilde U(uv, z)$, $V =  v \,\tilde V(uv, z)$, and $Z = \tilde Z(uv)$.  Symmetry under $U \leftrightarrow V$ and $u \leftrightarrow v$ then implies $\tilde U = \tilde V$.  Expanding $\tilde U = \tilde V$ and $\tilde Z$ in powers of $z$ and solving \req{gzuconstraint} order-by-order then gives a unique series solution which can be summed to yield:
\begin{equation}
\tilde U = \tilde V  = 1 - { 2 \, (1-u\, v) \, z^2 \over 4\, (1-u\, v)- u \, v \, z^2}, \ \ \
\tilde Z = { 4 \, (1-u\, v)^{3/2} \, z \over 4\, (1-u\, v) - u \, v \, z^2}.
\label{resummedcx}
\end{equation}	
Applying  \req{resummedcx} and \req{uvrt} to \AdS{3} in the form \req{AdS3}, one arrives at the metric
\begin{eqnarray}
ds^2
=
&\frac{1}{z^2}&  \left( -f(r,z)\, dt^2 + g(r,z)\, dr^2 + dz^2\right), \cr
{\rm where} \ \ \
f(r,z) &=&  \tanh^2 r \,\left(1 - z^2 \;\frac{\cosh^2 r + 1}{4 \, \cosh^2 r}\right)^{\! \! 2}, \cr
{\rm and} \ \ \
g(r,z) &=&  \left(1 + z^2 \;\frac{ \cosh^2 r - 3}{4 \, \cosh^2 r}\right)^{\! \! 2}.
\label{ads3fun}
\end{eqnarray}	
In this form, the solution is simply a quartic polynomial\footnote{As such, (\ref{ads3fun}) can be found by solving the Einstein equations order by order in $z$ for a static ansatz.  The series truncates at order $z^4$.   One finds a single arbitrary parameter whose value is fixed by smoothness, which requires the temperature of the bulk horizon to match that of the boundary black hole.   A similar structure arises for any static two-dimensional boundary metric.}  in $z$.

\begin{figure}[h]
\begin{center}
\includegraphics[scale=1.2]{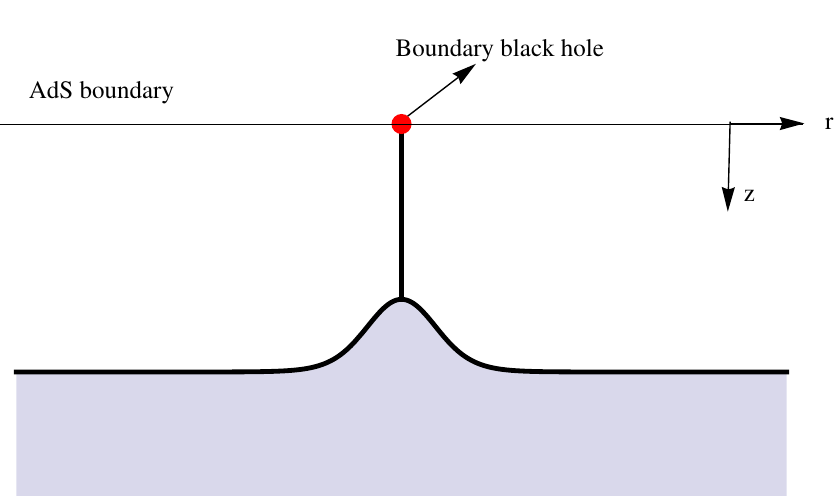}
 \caption{The plot of the bulk horizon for the black funnel geometry whose boundary is the two dimensional black hole \req{BF3formbdy}. We plot the location of the horizon in the $\{r,z\}$ coordinates. The shaded region lies inside the black hole as in \fig{f:fundrop}.}		
\label{f:2dfunnel}
\end{center}
\end{figure}

The horizon is located at the roots of $f(r,z) =0$:
\begin{equation}
z_H(r) = \frac{2 \, \cosh r}{\sqrt{\cosh^2 r +1}} \ , \;\; {\rm or} \;\; r = 0
\label{zH}
\end{equation}	
As a result, in these coordinates the horizon takes the shape of a funnel whose neck has collapsed, see \fig{f:2dfunnel}.
 One might worry that there is a double root of $f(r,z)$ at the point $r=0$, $z = z_H(r=0) = \sqrt{2}$.   However, the above analysis shows that this is an artifact of the coordinate transformation \req{resummedcx}, which is singular at $(r,z) = (0, \sqrt{2})$.  The full solution is merely AdS{}$_3$. In particular, the horizon is simply $UV=0$ and has constant temperature $T_H = \frac{1}{2 \pi}$.
For large $r$ the bulk horizon asymptotes to that of a planar BTZ black hole at this temperature. This follows from the fact that the horizon is at constant $z = 2$ as $r \to \infty$.

% Figure

It is straightforward to compute the boundary stress tensor from \req{ads3fun}, following  the strategy outlined in \cite{deHaro:2000xn}.   We find $T_{rt}=0$ and
\begin{equation}
T_{tt} = \frac{2  \,\ell_3}{16\pi\, G^{(3)}_N} \, \tanh^2 r \, \left(\frac{\cosh^2r -3}{2\, \cosh^2 r}\right)  \ , \qquad T_{rr} = \frac{2  \,\ell_3}{16\pi\, G^{(3)}_N} \, \frac{\cosh^2 r + 1}{2\, \cosh^2r},
\label{2dbhst}
\end{equation}	
which indeed takes the form of a perfect fluid at temperature $\frac{1}{2\pi}$ as $r \to \infty$. Note that the trace of the stress tensor \req{2dbhst} is non-zero at finite $r$ reflecting the conformal anomaly due to vacuum polarization effects. Other temperatures can be obtained by rescaling $u,v$ with corresponding results.  In the limit $T_H \to 0$ one finds $z_H \to \infty$, so that the planar part of the horizon recedes to the Poincar\'e horizon.  This is just what one expects from our general arguments for black funnels dual to  Hartle-Hawking states of 1+1 CFTs.  

While we have focussed on the particular boundary metric \req{BF3formbdy}, the general case displays similar features.  This follows from several facts noted above.  In particular, any static 1+1 asymptotically flat $T_H \neq 0$ black hole metric on the boundary i) is continuously related to \req{BF3formbdy} by a Weyl transformation, ii) has a unique smooth static stress tensor \cite{Christensen:1977jc}, and thus iii) determines a unique smooth \AdS{3} solution in Fefferman-Graham coordinates (which is in fact a 4th order polynomial in $z$).  As a result, this new solution will be related to \req{ads3fun} by a so-called Penrose-Brown-Henneaux transformation \cite{Penrose:1986ca,Brown:1986nw} (see e.g.\ the discussion in \cite{Papadimitriou:2005ii}).   The $T_H \to 0$ limits are also related in this way.

%~~~~~~~~~~~~~~~~~~~~~~~~~~~~~~~~~~~~~~~~~~~~~~~
\section{Funnels and Droplets from the \AdS{4} C-metric}
\label{s:adscmet}
%~~~~~~~~~~~~~~~~~~~~~~~~~~~~~~~~~~~~~~~~~~~~~~

Having completed our study of \AdS{3} black funnels, we now consider \AdS{4} solutions dual to states of $2+1$-dimensional CFTs on black hole backgrounds.  Since the boundary metric need not satisfy any particular equations of motion, the lack of Ricci-flat 2+1 black holes will not be a hindrance.  Below, we focus on boundary metrics that arise naturally in a family of exact bulk solutions known as the \AdS{4} C-metrics.  This family has previously been used to construct localized black holes on UV branes \cite{Emparan:1999wa, Emparan:1999fd} and also more recently to construct plasma ball solutions on IR branes \cite{Emparan:2009dj}.   In contrast, we study smooth solutions in which no branes are present.

The family of \AdS{4} C-metrics will provide examples both of black funnels and of black droplets suspended above a disconnected horizon.  The black funnel geometries are candidate duals to field theory Hartle-Hawking states. However, black droplets in the C-metric are generically suspended above a second horizon having a different temperature $T \neq T_H$.  While these solutions are static, this is an artifact of the bulk classical limit (or the large $N, \lambda$ limit from the field theory perspective).  Such solutions do not represent true equilibria and cannot be dual to Hartle-Hawking states.  Nevertheless, they  support our basic picture of black droplets and suggest that equilbrium configurations might be found in some more general family of solutions.

Below, we concentrate on certain critical values of the parameters for which the C-metric spacetime splits into two disconnected pieces separated by a new infinity, associated with new asymptotic regions.   On the boundary, the metric in the asymptotic region is ${\mathbb R}  \times \H^2$.  Roughly speaking, we will find that one piece contains a black funnel, while the other piece contains a  black droplet suspended above a deformed version of the black holes described in \cite{Emparan:1999fd}.  Note that the black holes of \cite{Emparan:1999fd} are the analogues of planar black holes for ${\mathbb R}  \times \H^2$ boundaries.   For more general values of the parameters, the asymptotic region disappears and the two pieces merge, making the solution difficult to interpret.  This general case will be discussed separately in \cite{Cmet}.     We describe the geometry of the \AdS{4} C-metric for the desired critical cases containing asymptotic regions in  \sec{s:geocmet} and then identify black funnels and droplets in \sec{s:bfcmet}.  

%~~~~~~~~~~~~~~~~~~~~~~~~~~~~~~~~~~~~~~~~~~~~~~~
\subsection{The geometry of the critical AdS C-metric}
\label{s:geocmet}
%~~~~~~~~~~~~~~~~~~~~~~~~~~~~~~~~~~~~~~~~~~~~~~

The C-metric solution in four dimensions corresponds physically to a pair of black holes being uniformly accelerated by a cosmic string. The most general form was found in \cite{Plebanski:1976gy} as a solution to Einstein-Maxwell theory with a cosmological constant.  This solution is specified by seven parameters, corresponding to the mass, angular momentum, an acceleration parameter, electric and magnetic charges, cosmological constant and a NUT parameter.

When the angular momentum and electric, magnetic, and NUT charges all vanish, the C-metric takes the form \cite{Emparan:1999fd}
\begin{equation}
ds^2 = \frac{\ell^2}{(x-y)^2} \, \left( -\frac{F(y)}{1+\lambda}\, dt^2 + \frac{dy^2}{F(y)} + \frac{dx^2}{G(x) } + G(x)\, d\phi^2\right) ,
\label{adsc}
\end{equation}	
\begin{equation}
{\rm where} \ \ F(\xi) = \lambda +\kappa\, \xi^2 +  2\,\mu\, \xi^3  , \ \ {\rm and} \ \  \qquad G(\xi)  = \lambda + 1 - F(\xi) = 1 -\kappa\, \xi^2 - 2 \,\mu \, \xi^3 \ .
\label{FGdefs}
\end{equation}	
In \req{adsc} we have rescaled $t$ by a constant relative to the form of the C-metric used in \cite{Emparan:1999fd}.  This metric satisfies Einstein's equations with negative cosmological constant,
\begin{equation}
\CE_{\mu\nu} = R_{\mu \nu} + \frac{3}{\ell_{4}^2} \, g_{\mu \nu} = 0 \ ,    \ \ \ {\rm where}  \ \ \
\ell_4 = \frac{\ell}{\sqrt{\lambda +1}}.
\label{ads4scale}
\end{equation}	
The limit $\lambda \to -1$ is thus Ricci flat and $\lambda <-1$ would give de Sitter-C metrics.
We consider only the asymptotically AdS case, $\lambda > -1$, for which the AdS boundary  is the surface $x=y$.  (A particular scaling limit with $\lambda \to -1$ will be considered in \App{scale}.)

In the metric \req{adsc}, $\ell$ captures the (inverse) acceleration, while $\lambda$ is related to the cosmological constant and $\mu \ge 0$ is the mass parameter of the black hole(s).  Finally, $\kappa$ is a discrete variable taking values $\pm1, 0$ and corresponds to different allowed topologies for the black holes; $\kappa =1$ corresponds to topologically spherical horizons while $\kappa = -1, 0$ corresponds to non-compact horizons with $\R^2 $ topology. For $\kappa =1$ one recovers the pure \AdS{4} spacetime when $\mu =0$. Likewise by taking a suitable $\ell^{-1} \to 0$ limit one can recover the standard \SAdS{4} black hole for $\kappa =1$.  For $\kappa =-1$ one obtains the topological black hole of \cite{Emparan:1999gf} and one can get the planar AdS black hole in the case when $\lambda =0$. A detailed discussion of the AdS C-metric properties can be found in \cite{Emparan:1999wa,Dias:2002mi} for $\kappa =1$ and in \cite{Emparan:1999fd} for other values of $\kappa$ (see also \cite{Cmet}).

We are, of course, particularly interested in the boundary metric, since we wish to restrict attention to geometries with a boundary black hole.  By choosing the conformal factor $(x-y)/\ell$ , we may take the boundary metric to be
\begin{equation}
ds^2_{\text{bdy}} = \gamma_{\mu\nu}\, dx^\mu \, dx^\nu =-\frac{F(x)}{1+\lambda}\, dt^2 + \frac{(1+\lambda)\, dx^2}{F(x)\, G(x)} + G(x)\, d\phi^2.
\label{indbdymet}
\end{equation}	
  We focus on cases where Hawking radiation can be cleanly separated from e.g., vacuum polarization effects due to spacetime curvature near the black hole.  In particular, we seek boundary metrics having a well-defined asymptotic region in which $\partial_t$ has constant non-zero norm but spatial distances become arbitrarily large.  

\begin{figure}[tp]
\begin{center}
\includegraphics[scale=1]{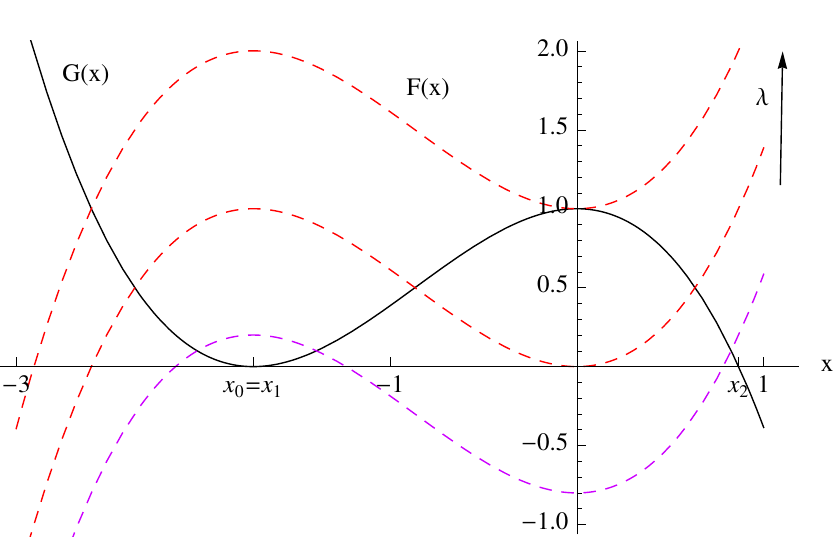}
  \caption{The functions $F(x)$ and $G(x)$ are plotted for $\mu = \mu_c$.   The solid curve is the function $G(x)$ while the dashed curves correspond to $F(x)$ plotted for various values of $\lambda$. The direction of increasing $\lambda$ is indicated in the figure.  The middle curve is $\lambda=0$. }
 \label{f:fgplots1C}
\end{center}
\end{figure}

To classify possible asymptotic regions, note that $x \to +\infty$ has $G(x) < 0$ and thus violates Lorentz signature.  Similarly $F(x) < 0$ as $x \to -\infty$, so that this region is not static, but instead lies inside a horizon.    Furthermore, zeros of $F(x)$ describe horizons where $\partial_t$ becomes null.  As a result, the desired asymptotic regions can arise only at roots of $G(x)$.  Since $F(x) = 1 + \lambda -G(x)$ and $\lambda > -1$, single roots $x_i$ of $G(x)$ lie at finite distance from the horizons.  Triple roots of $G(x)$ cannot arise from \req{FGdefs}, so we consider only the case where $G(x)$ has a double root $x_0=x_1$.  Near this root $||\partial_t|| \to - 1$ and the spatial part of the metric reduces to
\begin{equation}
ds_2^2 = \frac{dx^2}{(x-x_0)^2 } +  (x-x_0)^2 \, d\phi^2,
\label{x0double}
\end{equation}	
which is a metric on the Euclidean hyperboloid $\H^2$. This situation occurs only for
$\kappa =1$ and $\mu = \mu_c = \frac{1}{3\,\sqrt{3}}$, in which case the roots are $x_0 = x_1 = -\sqrt{3}$ and $x_2 = \sqrt{3}/2$.  We restrict attention to this case below, so that (since we hold $\ell_4$ in \req{ads4scale} fixed) the only parameter left to vary is $\lambda > -1$.\footnote{Of course, one can introduce an additional asymptotic region at some $x_{new}$ by changing the conformal frame.  However,  this process sends either $||\partial_t||$ or $||\partial_\phi||$ (or both) to infinity, so that the new asymptotic region is not of the desired form.} It is worth recording that  at $\mu = \mu_c$  we have $G'(x_0) =0$ and  $G''(x_0) =2 $, and that the corresponding derivatives for $F(\xi)$ can be determined by using the algebraic constraint $F(\xi) + G(\xi) = 1+ \lambda$.  Plots of the functions $F$ and $G$ for $\mu = \mu_c$ are shown in \fig{f:fgplots1C}.

% Figure
Recall that we seek solutions that, from a field theory point of view,  describe a deconfined plasma which permeates the asymptotic region.   For the AdS C-metric, this will be the region near $x = x_0$, where the boundary metric approaches that of ${\mathbb R}  \times \H^2$.
Fortuitously, AdS solutions dual to deconfined plasmas on $\R \times \H^2$ were studied in \cite{Emparan:1999gf}, where they were shown to minimize the free energy for all $T > 0$.  In \cite{Emparan:1999gf}, these solutions were called ``topological AdS black holes'', but we prefer the term ``hyperbolic black holes" which we use below.    The solution is
\begin{equation}
ds^2 = -f(\rho) \, dt^2 + \frac{d\rho^2}{f(\rho)} + \rho^2 \, d\Sigma_2^2,
\label{topads4}
\end{equation}	
where
\begin{equation}
f(\rho) =\frac{\rho^2}{\ell_4^2} -1-\frac{M}{\rho},
\label{ftopads}
\end{equation}	
and $d\Sigma_2^2$ is the metric on a unit $\H^2$.  The mass parameter $M$ must satisfy $M \ge  - \frac{2}{3\sqrt{3}} \ell_4$ and saturating the bound yields a smooth extremal black hole.

To compare the full AdS C-metric \req{adsc} (for $\kappa =1$, $\mu = \mu_c$)  with \req{topads4}, it is useful to consider the scaling limit
\begin{equation}
 \qquad \epsilon \to 0 \qquad {\rm with} \qquad
 X =\frac{x - x_0}{\epsilon}\ , \quad \Phi = \epsilon\, \phi ,\;\; t  , \;y \; \quad {\rm fixed},
\label{adscprops}
\end{equation}	
which zooms in on the asymptotic region.  This limit reduces the boundary metric to \req{x0double} (i.e., to $\H^2 \times {\mathbb R}$), while the bulk metric becomes
\begin{equation}
ds^2 = \frac{\ell^2}{(x_0 -y)^2} \, \left( -\frac{F(y)}{1+\lambda} \, dt^2 + \frac{dy^2}{F(y)} + \frac{dX^2}{X^2} + X^2\, d\Phi^2 \right).
\label{topads4a}
\end{equation}	
Here we used the fact that $G''(x_0) = 2 $.  The metric \req{topads4a} is just \req{topads4} in disguise. To recast \req{topads4a} into the form \req{topads4} we define
\begin{equation}
\rho \equiv \frac{\ell}{(x_0 - y)}.
\label{rhoequiv}
\end{equation}	
A calculation then yields
\begin{equation}
\frac{\ell^2\, F(y) }{(x_0 -y)^2} = (1+\lambda)\,\ell_4^2\, f(\rho), \ \ \ {\rm and} \ \ \  \frac{\ell^2 dy^2 }{(x-y)^2 F(y)} = \frac{d \rho^2}{f(\rho)},
\label{ell2Fy}
\end{equation}	
where we have identified
\begin{equation}
M = 2\, \ell_4 \, \mu_c\, \sqrt{1+\lambda} \ .
\label{Mmu}
\end{equation}	
This confirms the equality of \req{topads4a}  and \req{topads4}. Thus the solutions we obtain from the C-metric asymptote in an appropriate fashion to hyperbolic AdS black holes.

%~~~~~~~~~~~~~~~~~~~~~~~~~~~~~~~~~~~~~~~~~~~~~~
\subsection{Black funnels and droplets with asymptotically $\R \times \H^2$ boundary metric}
\label{s:bfcmet}
%~~~~~~~~~~~~~~~~~~~~~~~~~~~~~~~~~~~~~~~~~~~~~~

We  now have the pieces in place to identify the presence of any black-funnel- or black-droplet-like solutions.   The only parameter from \req{adsc} left to vary is $\lambda$, which controls both the location and the temperature of the horizons.  However, we have not yet specified the ranges of various coordinates.  To this end we introduce $z = x-y$, which plays the role of a radial coordinate.  The AdS boundary lies at $z=0$, so choosing either $z > 0$ or $z < 0$ yields an asymptotically AdS spacetime. In this sense our solutions come in pairs having the same boundary metric (\ref{indbdymet}).    

\begin{figure}[tp]
\begin{center}
\includegraphics[scale=0.9]{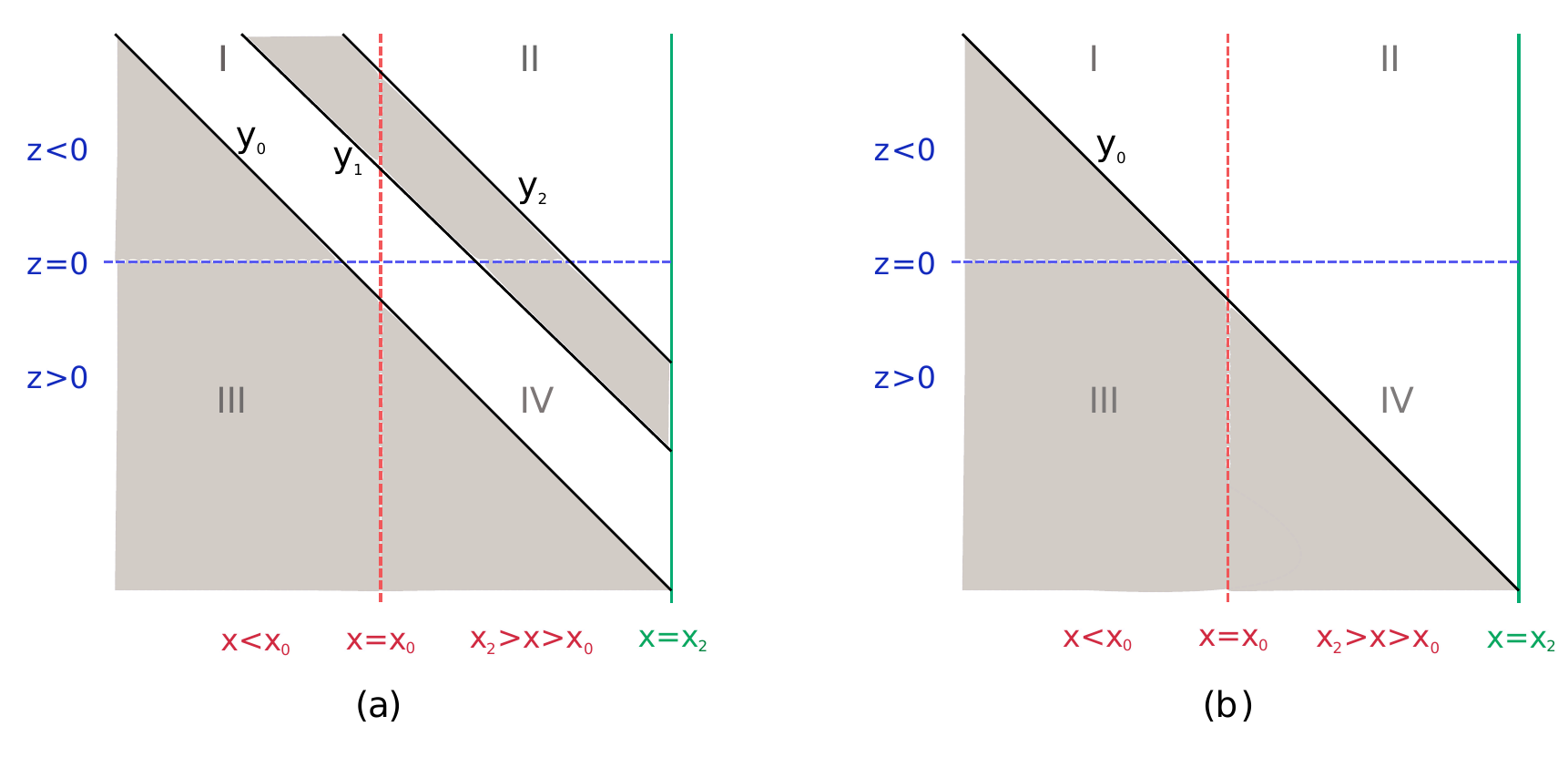}
  \caption{Horizons (diagonal lines) are plotted in the $x,z$ plane.  Note that $z$ increases downward while $x$ increases to the right.  Dotted lines denote infinities at $z=0$ and $x=x_0$.  As a result, each quadrant may be considered a complete spacetime unto itself. The left panel (a) corresponds to the situation when $-1 < \lambda < 0$ where $F(y)$ has three real roots and the right panel (b) corresponds to $\lambda >0$; compare with \fig{f:fgplots1C}.
  Note that in either case a conical singularity at $x = x_2$ will be visible from the point $x=x_0, z=0$ in quadrant IV.  }
\label{f:Horizons}
\end{center}
\end{figure}

In the $x$ direction,  Lorentz signature (non-negativity of $G(x)$) requires $x \le x_2$.  However, since the surface $x=x_0$ is infinitely far from other values of $x$, we can restrict attention to a range of $x$ which terminates at $x =x_0$, such that $G(x)$ stays positive.  This can be achieved by choosing either $x < x_0$ or $x_0 < x < x_2$; either choice yields a spacetime which is geodesically complete near $x = x_0$.    As a result, it is useful to refer to quadrants\footnote{The quadrants are defined with the origin in $x$ translated to $x=x_0$, i.e., 
quadrant I $=(z<0,x<x_0)$, quadrant II $=(z<0,x_0<x<x_2)$, etc.} in the $(x,z)$ plane as shown in figure \fig{f:Horizons}, below. The range of $\phi$ will be discussed later.

% Figure
\begin{figure}[tp]
\begin{center}
\includegraphics[scale=1]{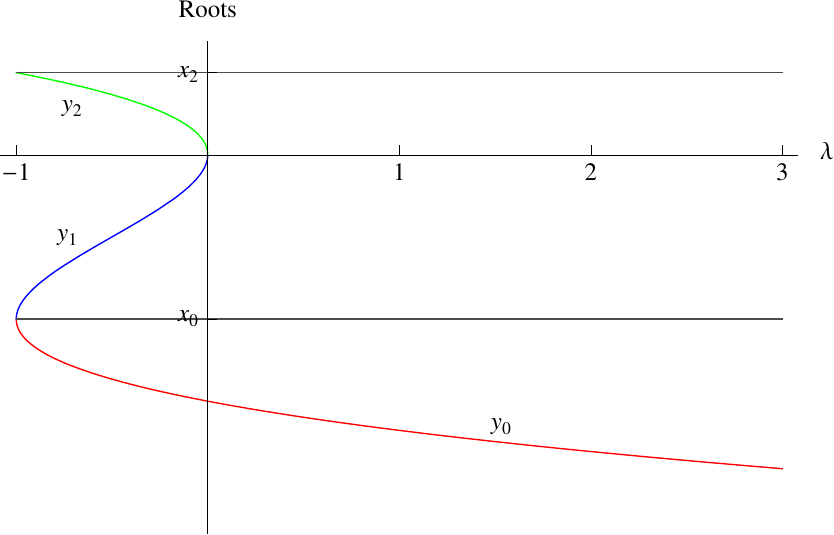}
  \caption{Real roots of $F$ (curved lines) are shown for  $\lambda > -1$ with $\mu =\mu_c$ and $\kappa =1$. The values of $x_0, x_2$ (straight lines) are shown for comparison.  At each $\lambda$, the smallest root $y_0$ satisfies $y_0 < x_0$.  All roots lie below $x_2$, and $y_0$ is the only real root for $\lambda > 0$.   }	
\label{f:Froots}
\end{center}
\end{figure}

It remains to interpret our solutions and to identify black funnels and black droplets.  Note that horizons occur on surfaces of fixed $y = x-z$ which lie at the roots of the cubic $F(y)$.  Depending on the value of $\lambda$, we can have either one real root ($y_0$) or three real roots  (ordered as $y_0 < y_1 < y_2$).  As shown in \fig{f:Froots}, for $\mu =\mu_c$ and $\kappa =1$, the smallest root $y_0$ satisfies $y_0 < x_0$, while $y_1,y_2 > x_0$ when they exist (for $-1 < \lambda \le 0$). We therefore have only the two situations which are sketched in \fig{f:Horizons}, except for various degenerate limits on which we comment below.

% Figure

\begin{figure}[tp]
\begin{center}
\includegraphics[scale=1]{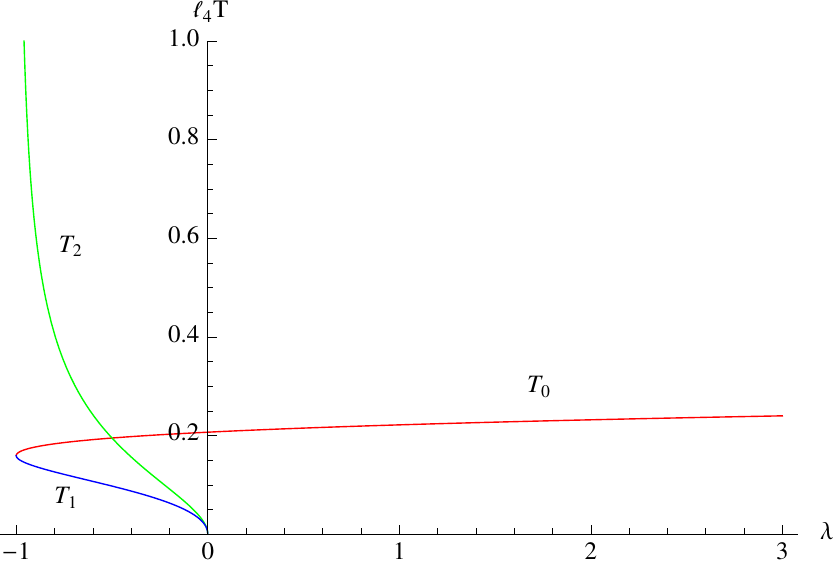}
  \caption{The temperatures for the various horizons $y= \{y_0,y_1,y_2\}$ as a function $\lambda$ for the $\kappa =1$, $\mu=\mu_c$ family of AdS C-metrics.  Note from \fig{f:Horizons} that the horizon at $y_2$ is always an inner horizon, so its temperature is not observable from outside the black hole. }\label{f:Tvslambda}
\end{center}
\end{figure}

We are concerned only with ``outer horizons,'' which by definition are those visible from the point $z=0$, $x = x_0$, i.e., from the asymptotic region of the boundary metric.  Such horizons are associated either with the root $y_0$ or with the root $y_1$.  A plot showing the temperature $T = \frac{|F'(y_i)|}{4\pi\, \sqrt{1+\lambda}}$ of each horizon is provided in \fig{f:Tvslambda}, but the temperature of any outer horizon near $x=x_0$ must agree with that of the hyperbolic black hole \req{topads4a}. Furthermore, as one may explicitly check, the agreement of the bulk metric near $x=x_0$ with the hyperbolic black hole \req{topads4a} implies that the boundary stress near $x=x_0$ also agrees with that of the hyperbolic black hole \req{topads4a}. So we indeed recover a thermal stress tensor in the asymptotic region (for a detailed discussion see \cite{Cmet}). Since we know from \cite{Emparan:1999gf} that this result describes the dual of the deconfined phase for any $T> 0$, we have candidate geometries describing the deconfined phase of the field theory on asymptotically $\R \times \H^2$ black holes as desired.

Due to this agreement, from the perspective of a given quadrant, an outer horizon is a black funnel if it extends continuously from the AdS boundary at $z=0$ through the bulk into the region near $x=x_0$; i.e.\ into the region that describes the   bulk dual of the field theory asymptotic region.  This is precisely the analogue of the statement from section \sec{heuristic} that, for asymptotically flat boundaries, a black funnel horizon will asymptote to the geometry of the planar AdS black hole.   From \fig{f:Horizons} we can see that this occurs for the $y_1$ horizon in quadrant II and for the $y_0$ horizon in quadrant III.

In contrast, we also encounter horizons that extend to the field theory asymptotic region $x=x_0$ in the bulk, but do not reach the AdS boundary at $z=0$.
These horizons are deformations of the hyperbolic black hole horizons \req{topads4} from \cite{Emparan:1999gf}.  This situation occurs for horizons at $y_1$ in quadrant I and $y_0$ in quadrant IV.  One can think of these horizons as being deformed in the bulk due to a combination of two effects: the acceleration provided by the cosmic string and the non-trivial boundary condition \req{indbdymet} at the AdS boundary.  

Finally, an outer horizon is a black droplet if it does not reach the field theory asymptopia $x=x_0$ in the bulk of the spacetime.  This occurs only for the $y_0$ horizon in quadrant I and for the $y_1$ horizon (when it exists) in quadrant IV.  As we discuss below, the quadrant I case is singular.  However, the quadrant IV case closes off smoothly in the bulk, satisfying our criterion for a black droplet. These horizons are analogs of the brane-world black holes of \cite{Emparan:1999wa, Emparan:1999fd} without the UV-brane (and in the specific region of parameter space $\kappa =1, \mu = \mu_c$).

 Before proceeding to take stock of the various solutions, let us note an interesting subtlety in our identification of the mass parameter $M$ for the hyperbolic black hole in \sec{s:geocmet}. If we take the coordinate region $z > 0$, then we have $\rho > 0$ from \req{rhoequiv}. Then the mass parameter is  indeed $M= 2\,\ell_4\, \mu_c\, \sqrt{1+ \lambda}$ as in \sec{s:geocmet}.  As a result, for $\lambda \to \infty$ we find $M \to \infty$ and thus $T \to \infty$ from \cite{Emparan:1999gf} (in particular, at large $\lambda$, $T \sim \lambda^{1/6}$). For the choice $z < 0$ however, \req{rhoequiv} would yield $\rho < 0$.  But $\rho \to -\rho$ in \req{ftopads} is equivalent to $M \to -M$.  So, for $z < 0$ we should identify $M = -2\, \ell_4 \, \mu_c \, \sqrt{1+ \lambda}$ as the mass parameter.

It remains only to discuss singularities.  The metric (\ref{adsc}) has curvature singularities at $x = \pm \infty$ and $y = \pm \infty$.  The singularity at $x  =  +\infty$ is irrelevant since we restrict to $x < x_2$.  In addition, the singularity at $y = -\infty$ is always shielded by a horizon.  However, the singularity at $y = + \infty$ is shielded by a horizon only for $\lambda \le 0$.  For $\lambda > 0$ it yields a naked singularity in quadrants I and II.  In particular, in the region near $x_0$ it coincides with the naked singularity of (\ref{indbdymet}) for $M < - \frac{2}{3 \sqrt{3}}\ell_4$.   In addition, the singularity at $x = -\infty$ is visible in quadrant I; e.g., along a curve of constant $y$. 

There is also a potential conical singularity associated with the fact that the $\phi$ circle degenerates only at $x=x_2$.   Smoothness at $x_2$ would imply that $\phi$ is periodic with period $\Delta \phi = \left| \frac{4\pi}{G'(x_2)} \right| =  \frac{8\pi \sqrt{3}}{9} $.  However, this is relevant only in quadrants II and IV.  Furthermore, we are interested in quadrant II only for $\lambda \le 0$, in which case the potential conical singularity at $x=x_2$ is shielded by a horizon.  Thus, at least for the interesting cases of quadrants I,II, and III, we are free to take $\phi$ to range either over  $\mathbb{R}$ or over a circle of any size.  However, in quadrant IV the potential singularity is visible from $z=0,x=x_0$ even when horizons are present and we must take $\Delta \phi =  \frac{8\pi \sqrt{3}}{9}$.

In summary, for the AdS C-metric solutions \req{adsc} with $\mu = \mu_c$, $\kappa =1$ and $\lambda > -1$ we encounter the following possibilites:
\begin{itemize}
\item {\bf Quadrant I:} There is a naked singularity at $x = -\infty$.  However, the horizon at $y_0$ may be thought of as a singular black droplet and, for $\lambda < 0$, $y_1$ is a (singular) deformed hyperbolic black hole with $0> M = -2\,\ell_4\, \mu_c \,\sqrt{1+ \lambda} \ge - \frac{2}{3 \sqrt{3} } \,\ell_4$.  For $\lambda =0$ this horizon is degenerate (the roots $y_1$ and $y_2$ merge), and asymptotes to the extreme hyperbolic black hole.  It disappears for $\lambda > 0$, revealing a second curvature singularity at $y = +\infty$.

\item {\bf Quadrant II:} For $\lambda < 0$, $y_1$ is a black funnel.  It asymptotes to a hyperbolic black hole with $0> M = -2\,\ell_4\, \mu_c \,\sqrt{1+ \lambda} \ge - \frac{2}{3 \sqrt{3} }\, \ell_4$.  For $\lambda =0$ this horizon is degenerate, and asymptotes to the extreme hyperbolic black hole.  No singularities are visible for $\lambda \le 0$, but the horizon disappears for $\lambda > 0$, revealing a naked curvature singularity and a potential conical singularity.

\item {\bf Quadrant III:} The horizon at $y_0$ is a black funnel.   It asymptotes to a hyperbolic black hole with $M = 2\,\ell_4 \,\mu_c \,\sqrt{1+ \lambda} > 0$.  No singularities are visible.

\item {\bf Quadrant IV:} The horizon at $y_0$ is a deformed hyperbolic black hole with $M = 2\ell_4 \mu_c \sqrt{1+ \lambda} > 0$.   For $\lambda < 0$, the horizon at $y_1$ is a black droplet.  It becomes degenerate at $\lambda =0$ and disappears for $\lambda > 0$.   The potential conical singularity at $x=x_2$ is visible for all $\lambda$, but taking $\Delta \phi =  \frac{8\pi \sqrt{3}}{9}$ yields a spacetime with no visible singularities.
\end{itemize}

The above black funnels are candidate duals of field theory Hartle-Hawking states as discussed in \sec{heuristic}.  However, the black droplet solutions are not, since they are accompanied by a hyperbolic black hole with a different temperature $T \neq T_H$; see \fig{f:Tvslambda}.  Nonetheless, these solutions support the general picture of droplets discussed in \sec{heuristic} and suggest that equilibrium solutions may exist within some larger family of solutions.

We conclude with a brief discussion of issues associated with taking the range of $\phi$ to be compact.   In this case, the boundary spacetime near $x=x_0$ is a quotient of ${\mathbb R}  \times \H^2$.  This quotient is non-compact, but the metric \req{x0double} shows that any region near $x=x_0$ has finite volume.
Now,  conformal field theories on hyperbolic manifolds with finite volume are prone to instabilities, as conformally coupled scalars will pick up negative masses.   In the gravity description the tunneling amplitude may be easily computed by looking for the nucleation probability of M2-branes \cite{Seiberg:1999xz}.   If  \req{x0double} were the full boundary metric, this would not be of concern as the field theory would live on a space of infinite volume even when $\phi$ has finite range,   suppressing the rate for the scalars to tunnel out to infinity from a local minimum.  However, a black hole horizon in \req{indbdymet} cuts off this divergence, so that the field theory lives on a spacetime for which the volume outside the horizon is finite.  We therefore expect our \AdS{4} solutions to be unstable to bulk quantum processes (boundary $1/N$ effects) when the range of $\phi$ is compact.  It would be interesting to understand if this feature is associated in an essential way with the properties of quadrant IV, and in particular  with  black droplets suspended above $M > 0$ hyperbolic black holes.

%~~~~~~~~~~~~~~~~~~~~~~~~~~~~~~~~~~~~~~~~~~~~~~
\section{Discussion}
\label{s:discuss}
%~~~~~~~~~~~~~~~~~~~~~~~~~~~~~~~~~~~~~~~~~~~~~~

Motivated by the desire to better understand Hawking radiation in strongly coupled field theories, we have analyzed a dual formulation of the problem using the AdS/CFT correspondence.  The quantum dynamics of the field theory  maps via this correspondence to the classical dynamics of Einstein's equations in a bulk AdS spacetime.
Given a field theory in a black hole background $\CB_d$ we are instructed to find all the negatively curved Einstein manifolds $\CM_{d+1}$ whose boundary is $\CB_d$.

We have two distinct possible duals for the Hartle-Hawking state of the field theory in the boundary black hole background:
\begin{itemize}
\item The black funnel, which is a  bulk solution with a single connected horizon.
\item A black droplet and a planar or hyperbolic AdS black hole, i.e., a solution with two disconnected horizons.
\end{itemize}

A black hole spacetime $\CB_d$ typically has a horizon size $R$ and a temperature $T_H$ and, as we have argued in \sec{heuristic}, general considerations indicate that at least for asymptotically flat boundary black holes the black funnel is preferred for $R \, T_H \gg 1$ while the multiple-horizon solution dominates for $R \, T_H \ll 1$. In either case the bulk solution is dual to the deconfined phase of the field theory in the boundary black hole background, i.e., the free energy and the quantum stress tensor of the field theory scales like the central charge $c$ of the CFT.  From the field theory point of view, which possibility occurs depends on the rate at which the field theory plasma equilibrates with the (non-dynamical) field theory black hole, a concept we refer to as the ``coupling'' of plasma excitations to the black hole in the sense that strong coupling means fast equilibration.   We interpreted the weak coupling case as arising when the plasma excitations have a typical physical size $R_e$ which is greater than the size $R$ of the black hole.  The size of such excitations is an analogue of the gravitational form factors discussed in \cite{Polchinski:2001ju}.  

While pointlike particles display no sharp transition in the strength of their coupling to black holes, our arguments indicate a new kind of phase transition in the large  $N$, strong coupling ($\lambda \gg 1$) limit.  Due to the change in connectivity of the bulk horizon, the on-shell action in the bulk will be non-analytic in parameters that control the size of the boundary black hole. This then implies that the partition function of the field theory in the boundary black hole background will exhibit non-analytic behavior characteristic of phase transitions. It would be very interesting to probe this transition, and in particular the regime $R\,T_H \sim 1$ which is relevant for the field theory in an asymptotically flat Schwarzschild black hole background.  It would be natural for this transition to be similar to the Gregory-Laflamme transition, and to display corresponding singular solutions with ``pinched horizons"  connecting parts of the phase diagram with different topologies.

We focussed here on constructing simple analytic examples of funnels and droplets in low dimensions.  We showed that only funnels can arise for 1+1 boundary black holes, and constructed a corresponding \AdS{3} solution in \sec{s:bfunnelbtz}. We then explored the AdS C-metric family of spacetimes in \sec{s:adscmet} to find examples of funnels and droplets for 2+1 boundary black holes.  Both types of solutions were found, but only for boundary black holes that are asymptotic to $\mathbb{R} \times \H^2$.  Asymptotically flat boundary black holes simply do not arise in this family of spacetimes.  Also, due to the peculiarities of the C-metric family, the droplet solutions did not describe true equilibrium configurations.  Instead, the droplets were suspended above hyperbolic black holes of different temperatures $T \neq T_H$.  Nevertheless, these droplet solutions support the general picture described in \sec{heuristic} and suggest that equilibrium droplets may exist in some larger class of solutions.

It would be interesting to compare the \AdS{4} C-metric solutions of \sec{s:adscmet} in detail with the predictions of \sec{hBH} for asymptotically flat boundary black holes.  However, several factors make this difficult:  i) there is an extra scale from the curvature of the $\H^2$, ii) our droplet solutions are not in equilibrium, and iii) we have at hand only those solutions  that belong to the C-metric family; more general solutions will certainly exist.  As a result, we content ourselves with the following observations:

\begin{itemize}

\item The only outer horizon whose temperature can be large is the $y_0$ horizon, which is a smooth horizon only in quadrants III and IV of \fig{f:Horizons}.  High temperatures occur at large values of the parameter $\lambda$, for which quadrant III contains a black funnel and quadrant IV contains a hyperbolic black hole which does not reach the boundary.\footnote{For $\lambda > 0$ the quadrant IV boundary metric has no horizon and contains no black hole.}  Thus, while we find high temperature black funnels, as predicted by \sec{heuristic} there are no droplets suspended above high temperature hyperbolic black holes.

\item The only $T_H \to 0$ limit of a smooth outer horizon occurs for the $y_1$ horizon in quadrant II for $\lambda \to 0$.  This horizon is a black funnel, and not the droplet predicted by \sec{heuristic}.  This may be related to the $\H^2$ curvature, which one expects to be relevant for small $T_H$.  It may also be that a droplet solution exists in some larger family of solutions, and that this droplet solution is thermodynamically preferred.

\item Limits where the outer horizons horizons approach $x_2$ might be thought of as small black holes, i.e., ``small R'' in the sense of \sec{heuristic}, but this situation doesn't arise within the AdS C-metric family of solutions.

\item One might also wish to study ``large R'' limits.  It might appear that $\lambda \to -1$ provides such a limit, where the $y_0$ and $y_1$ horizons approach $x_0$.    However, we must take this limit holding the bulk AdS scale $\ell_4$ fixed, so that $\ell \to 0$ and our metric \req{adsc} degenerates. A $\lambda \to -1$ scaling limit is described in \App{scale}.  This limit yields boundary black holes of finite size, and bulk solutions corresponding to either black funnels or (non-compact analogues of) black droplets suspended above hyperbolic black holes. This time, however, the droplets and hyperbolic black holes are in equilibrium.  The scaling limit takes an especially simple form (\ref{adsclm1m}).
\end{itemize}
In summary, while the AdS C-metric family of solutions yields interesting examples of black funnels and droplets, these unfortunately don't lend themselves to a detailed verification of the predictions of \sec{heuristic}. 

Our work raises a number of interesting questions and suggests many directions for further research. These include exploring the above phase transition and a more thorough exploration of the space of solutions (probably through numerical work) in both 2+1 and higher dimensions.   Some particularly interesting further avenues for exploration are as follows:

\noindent {\bf Confining field theories, plasma balls, black funnels, and black droplets:}  It is interesting to consider field theories with a first order deconfinement transition at some $T_D > 0$.  These may be addressed using AdS/CFT \cite{Witten:1998zw} by making a so-called Scherk-Schwarz compactification of the boundary field theory, for which $\CB_d$ asymptotes to $\Sp^1 \times \R^{d-2,1}$ with anti-periodic boundary conditions on any fermions along the $\Sp^1$.  The presence of a new scale in the problem, viz., the size of the $\Sp^1$ (call it $L$) ensures that one can have a proper phase transition as one varies the product $T\, L$. 

When $\CB_d$ is precisely $\Sp^1 \times \R^{d-2,1}$, the deconfined phase is again the planar black hole (\ref{planarBH}) while the confined phase is the so-called AdS soliton \cite{Horowitz:1998ha}.  The latter contains an ``infra-red (IR) floor" where the $\Sp^1$ smoothly caps off at a finite value of the AdS radial coordinate (or the Fefferman-Graham coordinate $z$) without generating a horizon, so that the bulk spacetime looks like a cigar geometry. This situation leads to interesting new solutions with compact horizons localized near the IR floor \cite{Aharony:2005bm}.  Such solutions are called ``plasma balls"  since they are dual to balls of deconfined plasma that nucleate from the confined state. They have a characteristic size which depends on how close the temperature is to the deconfinement temperature $T_D \sim 1/L$, and in fact their size diverges as the temperature approaches $T_D$. We expect that similar solutions should arise when $\CB_d$ contains a black hole.  In addition, this setting allows a simple smooth solution where the boundary metric is $\Sp^1$ times any Ricci-flat black hole (e.g., Schwarzschild), where the bulk horizon is a string that extends from the AdS boundary to the IR floor.\footnote{We thank Rob Myers for pointing this out to us.} This solution should probably be called the Schwarzschild-black string AdS soliton, but we will refer to it as the Schwarzschild soliton string.

One would like to explore the full phase diagram as one varies $R,T_H$ of the boundary black hole (keeping the size of the circle $L$ fixed) in order to understand how these Schwarzschild soliton strings are connected to plasma balls, black funnels, black droplets, and so forth.   One also wishes to understand which solutions are dynamically and/or thermodynamically stable.    In particular, the Schwarzschild soliton strings should  be dynamically unstable for $R \sim T^{-1}  \ll T_D^{-1}$.   One might expect that the instability leads to black droplets as originally conjectured by \cite{Chamblin:1999by} in the context of AdS black strings, but the possibility of a rather more spectacular instability to black funnel formation has not been ruled out.  Indeed, even if this does not occur for Schwarzschild soliton strings, it may occur for more general boundary metrics, providing contexts in which black holes in some sense ``melt'' confined phases dynamically and catalyze their transition to deconfined phases.

\noindent  {\bf Other field theory asymptotics:}
It is also interesting to discuss the case studied in \cite{Gregory:2008br} where the field theory itself lives in an asymptotically AdS$_d$ spacetime such as Schwarzschild-AdS$_d$ or BTZ for $d=3$.  In such cases, the bulk dual is an asymptotically AdS${}_{d+1}$ spacetime whose  boundary metric is asymptotically AdS$_{d}$.   This case appears to have additional subtleties (owing to the norm of the timelike Killing field growing on the boundary) and will be discussed separately in a forthcoming work \cite{toappear}.

\noindent  {\bf Rotation:} Rotating asymptotically flat field theory black holes do not admit well-defined Hartle-Hawking states \cite{Kay:1988mu} due to features associated with super-radiance.  From the bulk perspective, one expects that black funnels cannot exist for rotating boundary black holes, since it is not possible for the funnels to be rigidy rotating. As a result, in this work we have focussed on non-rotating field theory black holes. However, the rotating case will be explored in \cite{toappear}.

\noindent  {\bf Non-equilibrium solutions:}   Small disturbances of equilibrium solutions generally result in flows of heat.  We suggested interesting examples in \sec{heuristic}, based on placing black funnels, or droplets suspended above planar black holes, in finite volume.  In the funnel case, changing the volume should result in a large flow of heat along the horizon on short timescales.  In the droplet/planar case, the planar horizon will respond rapidly to the change in volume by changing its temperature.  It is only over very long timescales associated with Hawking radiation in the bulk that it will re-equilibrate with the droplet, whose temperature is fixed to be $T_H$ by its connection to the boundary black hole.  Though we consider only deconfined phases, in this latter case the heat exchange with the black hole would be more typical of what one would expect in a confined phase of the field theory.

The solutions from \sec{s:adscmet} containing black droplets suspended above hyperbolic black holes with a different temperature provide some support for this picture.  These solutions are static in the limit of a classical AdS bulk, though they will feature a flow of heat when bulk quantum effects are included.  It would be interesting to explore the space of such solutions more thoroughly, especially for asymptotically flat black holes.  

Black funnel solutions displaying a constant flow of heat either into or out of the boundary black hole may also prove tractable, at least with numerical methods.  One may be able to fix the temperature of the shoulder region of \fig{f:fundrop} independently of the boundary black hole temperature $T_H$.  This is a stationary situation, in which one expects a stationary bulk AdS solution.  However, rather remarkably, the event horizon cannot be a smooth Killing horizon, as such horizons have uniform temperature.  While stationary black holes without Killing horizons are unfamiliar, there is no contradiction with the known rigidity theorems (e.g.\ \cite{Hollands:2006rj}) which require a compactly generated horizon.   Furthermore, the constant flow of heat implies that the boundary stress tensor must diverge on the past horizon of the field theory black hole, suggesting that the bulk solution is also singular on the past horizon. The situation is reminiscent of the considerations of the fluid-gravity correspondence \cite{Bhattacharyya:2008jc} where arbitrary fluid flows in the field theory were mapped to inhomogeneous, dynamical black holes in the bulk AdS spacetime, albeit with a crucial distinction: since the fluid-gravity solutions are necessarily dynamical, no novel stationary solutions are known in that context.

The possibility of such novel solutions raises many further questions about the relation between the boundary heat flow and the bulk horizon generators, compatibility with Raychaudhuri's theorem, and the bulk description of the flows of negative energy across the horizon of the field theory black holes that one associates with Hawking radiation in the field theory.  After all, from the field theory perspective, it must be this Hawking radiation that carries heat outward from the field theory black hole.

\noindent  {\bf Implications for Braneworlds:}  In this work we have concentrated on bulk solutions which extend to the AdS boundary, where we can fix any metric as a boundary condition.  However, one may imagine attempting to cut off such solutions with a so-called ``ultra-violet (UV) brane'' so that the solution instead ends on a dynamical boundary at finite distance.  One expects this to be possible at least at the level of initial data, though because the UV brane is dynamical the resulting evolution may take a rather different form.  Nevertheless, it would be very interesting to understand the dynamics that result from cutting off any of the classes of solutions described above.  One expects equilibrium solutions to remain static, but not necessarily stable.  

A primary question in this field remains the end-state of the instability associated with the AdS black strings (or equivalently the Schwarzschild soliton strings from our discussion of confining theories).  While one expects a decay to a stable droplet attached to the brane, an instability to the formation of a black funnel is not ruled out.  If it arises, this would lead to a drastic modification of our understanding of brane-world black holes.

%_____________________________________________
\subsection*{Acknowledgements}
%_________________________________________________

It is a pleasure to thank Roberto Emparan, Jim Hartle, Gary Horowitz, Shamit Kachru, Hong Liu, Rob Myers,  Joe Polchinski, Ralf Sch\"utzhold and Toby Wiseman for very interesting discussions.  VEH and MR would like to thank the KITP for wonderful hospitality during the workshop ``Fundamental Aspects of Superstring Theory", as well as the  Pedro Pascual Benasque Center of Science and the Aspen Center for Physics for excellent hospitality during the course of this project. In addition, VEH, DM and MR would like to thank the ICTS, TIFR for hospitality during the Monsoon workshop in string theory where this project was initiated. VEH and MR are supported in part by STFC Rolling grant and by the National Science Foundation under the Grant No. NSF PHY05-51164.  DM was supported in part by the US National Science Foundation under Grant No. PHY05-55669, and by funds from the University of California.

\appendix

\section{The limit $\lambda \to - 1$ for $\mu = \mu_c$, $\kappa =1$}
\label{scale}

This appendix studies the limit $\lambda \to - 1$ for $\mu = \mu_c$, $\kappa =1$, holding the bulk AdS scale $\ell_4$ fixed.  Since $\ell \to 0$, the metric \req{adsc} degenerates.  In contrast, taking the scaling limit : %
\begin{eqnarray}
&&\lambda \to -1 , \;\;\mu = \mu_c  , \qquad \text{with} \nonumber \\
&&X = \frac{x-x_0}{\sqrt{\lambda +1}} \ , \quad Y = \frac{y-x_0}{\sqrt{\lambda +1}} \ , \quad \frac{\ell}{ \sqrt{1+\lambda }} \ , \quad \Phi = \sqrt{1+\lambda}\, \phi , \quad t, \quad \text{fixed}  ,
\label{morelimits}
\end{eqnarray}	
yields the finite result:
\begin{equation}
ds^2  = \frac{\ell_4^2}{(X-Y)^2} \, \left( -(1-Y^2)\,dt^2  + \frac{dY^2}{1-Y^2} +\frac{dX^2}{X^2} + X^2 \, d\Phi^2\right),
\label{adsclm1m}
\end{equation}	
with horizons at $Y = \pm 1$.  The induced metric on the boundary \req{indbdymet} in this limit reduces to
\begin{equation}
ds^2 = -(1-X^2)\, dt^2 +\frac{dX^2}{X^2\,(1-X^2)} + X^2\, d\Phi^2\ .
\label{indbdymetds} 
\end{equation}	

We see that the horizons in the boundary metric (\ref{indbdymetds}) are of finite size and finite temperature, $T_H = 1/2\pi$, so that they do not probe the large $R\,T_H$ regime discussed in \sec{heuristic}.  Nevertheles, quadrants II and III describe black funnels, while quadrants I and IV describe non-compact versions of black droplets\footnote{The droplet horizons now extend an infinite distance through the bulk to $X = \pm \infty$.} suspended above hyperbolic black holes. This time the droplets and hyperbolic black holes are in equilibrium.  All singularities have disappeared.  Furthermore, near $X=0$ the solution asymptotes to the $M=0$ hyperbolic black hole (\ref{topads4}), which is just \AdS{4} in accelerated coordinates \cite{Emparan:1999gf}.  The boundary stress tensor vanishes for this solution, and can be shown to vanish for \req{adsclm1m} as well.

%%%%%%%%%%%%%%%%%%%%%%%%%%%%%%%%%%%%%%%%%%%%
%\bibliography{bfunnel13}
%\bibliographystyle{utphys}
%%%%%%%%%%%%%%%%%%%%%%%%%%%%%%%%%%%%%%%%%%%%
\providecommand{\href}[2]{#2}\begingroup\raggedright\endgroup

\end{document}